\documentclass[twocolumn,showpacs,preprintnumbers,amsmath,amssymb,floatfix]{revtex4}
\usepackage{graphicx}
\usepackage{dcolumn}
\usepackage{bm}
\usepackage{amsthm}
\usepackage{amsmath}
\usepackage{amsfonts}
\usepackage{mathrsfs}

\newcommand{\be}{\begin{equation}}
\newcommand{\ee}{\end{equation}}
\newcommand{\ba}{\begin{eqnarray}}
\newcommand{\ea}{\end{eqnarray}}
\newcommand{\non}{\nonumber}
\newcommand{\bal}{\begin{align}}
\newcommand{\eal}{\end{align}}

\newcommand{\n}[1]{\label{#1}}

\newcommand{\eq}[1]{(\ref{#1})}

\newcommand{\htt}{\hat{t}}
\newcommand{\hr}{\hat{r}}
\newcommand{\hrho}{\hat{\rho}}
\newcommand{\hta}{\hat{\tau}}
\newcommand{\hal}{\hat{\alpha}}
\newcommand{\hbe}{\hat{\beta}}

\newcommand{\ha}{\hat{a}}
\newcommand{\hb}{\hat{b}}
\newcommand{\hc}{\hat{c}}
\newcommand{\hd}{\hat{d}}

\newcommand{\hhh}{\, \hspace{0.3cm}}
\newcommand{\hhhhh}{\, \hspace{0.01cm}}

\newenvironment{rcase}
{\left.\begin{aligned}}
{\end{aligned}\right\rbrace}

\begin{document}
\title{Analysis of the Fisher solution}

\author{Shohreh Abdolrahimi}
\email{sabdolra@phys.ualberta.ca}
\author{Andrey A. Shoom}
\email{ashoom@phys.ualberta.ca}
\affiliation{Theoretical Physics Institute, University of Alberta, Edmonton, Alberta,
Canada,  T6G 2G7}
\date{\today}
\begin{abstract}
We study the $d$-dimensional Fisher solution which represents a static, spherically symmetric, asymptotically flat spacetime with a massless scalar field. The solution has two parameters, the mass $M$ and the ``scalar charge" $\Sigma$. The Fisher solution has a naked curvature singularity which divides the spacetime manifold into two disconnected parts. The part which is asymptotically flat we call the {\em Fisher spacetime}, and another part we call the {\em Fisher universe}. The $d$-dimensional Schwarzschild-Tangherlini solution and the Fisher solution belong to the same theory and are dual to each other. The duality transformation acting in the parameter space $(M,\Sigma)$ maps the exterior region of the Schwarzschild-Tangherlini black hole into the Fisher spacetime which has a naked timelike singularity, and interior region of the black hole into the Fisher universe, which is an anisotropic expanding-contracting universe and which has two spacelike singularities representing its ``Big Bang" and ``Big Crunch". The Big Bang singularity and the singularity of the Fisher spacetime are {\em radially weak} in the sense that a 1-dimensional object moving along a timelike radial geodesic can arrive to the singularities intact. At the vicinity of the singularity the Fisher spacetime of nonzero mass has a region where its Misner-Sharp energy is negative. The Fisher universe has a marginally trapped surface corresponding to the state of its maximal expansion in the angular directions. These results and derived relations between geometric quantities of the Fisher spacetime, the Fisher universe, and the Schwarzschild-Tangherlini black hole may suggest that the massless scalar field transforms the black hole event horizon into the naked radially weak disjoint singularities of the Fisher spacetime and the Fisher universe which are ``dual to the horizon." 
\end{abstract}

\pacs{04.20.Dw, 04.20.Gz, 04.20.Jb, 04.50.Gh \hfill Alberta-Thy-18-09}
\maketitle

\section{Introduction}

In this paper we study a solution which was discovered by Fisher \cite{Fisher}. Later the solution was rediscovered by many authors (see, for example, \cite{Wym,Agn,Rob}) and usually referred to as the Janis-Newman-Winicour solution \cite{JNW}. Here we study the $d$-dimensional $(d\geqslant4)$ generalization of this solution which was given in \cite{Xan} . The solution represents a static, spherically symmetric, asymptotically flat spacetime with a massless scalar field. A massless scalar field is related to a massless particle of zero spin. Such particles are not known, and all known zero spin particles are massive. Thus, such a field may be not realistic (unless a zero spin massless particle is discovered). However, in some cases one may consider such a field as an approximation for a massive scalar field, or regard a massless scalar field as a toy model, which is often useful for its simplicity. There is a more serious reason to consider the Fisher solution as unphysical, for it represents a naked curvature singularity. 

The classical description of spacetime breaks down at a curvature singularity. However, spacetime singularities arise in a very large class of solutions of the general theory of relativity, and in fact in very reasonable physical conditions which respect causality and energy conditions \cite{Haw}. The trouble with naked singularities (except agreeably with the Big Bang one which is in our past) is that they are naked, i.e., one could potentially ``see a breakdown of physics" if a naked singularity is present. To avoid formation of a naked singularity in real physical processes, such as gravitational collapse, which are described by classical laws of the general theory of relativity, the cosmic censorship conjecture was formulated, first in weak \cite{Pencccw} and later in strong form \cite{Pencccs}. However, the present issue of its validity is very much open \cite{Penccc}. 

In attempts to test cosmic censorship, many models of gravitational collapse were studied analytically and numerically (for a popular survey of the subject see \cite{Joshi}).  It was found that in certain conditions naked singularities do form. For example, they may form as a result of collapse of collisionless gas spheres \cite{Shap}, or self-similar collapse of a massless, minimally coupled scalar field where the second type phase transition from black hole to naked singularity takes place \cite{Brad}. However, such examples should be considered with caution, for a rigorous analysis may suggest that the detected naked singularity formation may be ambiguous \cite{Wald}. A review \cite{Har} has many other examples as well as discussion of gravitational radiation and quantum particle creation by naked singularities. There is a recent proposal to search for a naked singularity using Kerr lensing \cite{Wer}. These examples may imply that we have to study naked singularities rather than disregard them.    

Here we study the naked singularity of the Fisher solution which is due to a massless scalar field. The reasons for such a study is to understand deeper how such a field affects spacetime and what type of singularity it ``produces." For example, it was shown that a massless scalar field ``converts" the Cauchy horizon of a Kerr-Newman black hole into a strong curvature singularity \cite{Mustafa}. Another example is a weak instantaneous curvature singularity which appears at the moment of a wormhole formation when a ghost massless scalar field is present \cite{Hideki}. On the other side, it was shown that quantum effects may prevent the formation of a naked singularity due to gravitational collapse of a homogeneous scalar field \cite{Ritu}. This may suggest that a curvature singularity due to massless scalar field may be ``smoothed out" by quantum effects. 

The main idea of our study is to analyze the naked curvature singularity of the Fisher solution and to show that indeed, a spacetime curvature singularity (at least in our example) may be a complex object and should be scrutinized carefully.  
 
This paper is organized as follows. In Sec. II we present the $d$-dimensional Fisher solution and discuss its general properties. In Sec. III we study curvature singularities of the Fisher solution. Causal properties of the Fisher solution are discussed in Sec. IV. In Sec. V we present an isometric embedding of the Fisher solution. Using results of the previous sections, we return to a discussion of the Fisher solution in Sec. VI. Section VII contains a summary and discussion of our results. Additional details illustrating  our calculations are given in the appendixes. In this paper we set $G_{(d)}=c=1$, where $G_{(d)}$ is the $d$-dimensional $(d\geqslant 4)$ gravitational constant. The spacetime signature is $+(d-2)$. We use the notations and conventions adopted in \cite{MTW}.

\section{The Fisher solution}

\subsection{Metric}

Let us present a $d$-dimensional generalization of the Fisher solution, which is static, spherically symmetric, asymptotically flat spacetime with a massless, minimally coupled scalar field. The corresponding action has the following form:
\be\n{2.1}
\mathcal{S}[g_{ab},\,\varphi]=\frac{1}{16\pi}\int d^dx\sqrt{-g}\left(R-\tfrac{d-2}{d-3}g^{ab}\varphi_{,a}\varphi_{,b} \right)\,,
\ee
where $R$ is the $d$-dimensional Ricci scalar and $\varphi$ is the massless, minimally coupled scalar field. Here and in what follows $(...)_{,a}$ stands for the partial derivative of the expression $(...)$ with respect to the coordinate $x^a$.

The energy-momentum tensor of the scalar field is
\be\n{emt}
T_{ab}=\frac{1}{8\pi}\frac{d-2}{d-3}\left(\varphi_{,a}\varphi_{,b}-\frac{1}{2}g_{ab}\varphi_{,c}\varphi^{,c}\right)\,.
\ee
Thus, the corresponding Einstein equations are
\be\n{2.2}
R_{ab}= \tfrac{d-2}{d-3}\varphi_{,a} \varphi_{,b}\,.
\ee
The scalar field solves the massless Klein-Gordon equation
\be\n{2.3}
\nabla^a\nabla_a\varphi=\frac{1}{\sqrt{-g}}\left(\sqrt{-g}g^{ab}\varphi_{,a}\right)_{,b}=0\,.
\ee
Here $\nabla_a$ stands for the covariant derivative defined with respect to the $d$-dimensional metric $g_{ab}$. 
An explicit form of Eqs. \eq{2.2} and \eq{2.3} for a static, spherically symmetric spacetime is given in Appendix A. A static, asymptotically flat, spherically symmetric solution to Eqs. \eq{2.2} and \eq{2.3} was derived in \cite{Xan} in isotropic coordinates, which bring the Einstein equations into a form more suitable for integration. Here we present the solution in different (Schwarzschild-like) coordinates \cite{Iso}. Duality transformation presented in the next subsection allows one to derive this solution without integration of the Einstein equations. The Fisher metric reads
\ba\n{2.4}
ds^2=-F^{S}dt^2+F^{\tfrac{\mathstrut 1-S}{d-3}-1}dr^2+r^2F^{\tfrac{\mathstrut1-S}{d-3}}d\Omega^2_{(d-2)}\,,
\ea
where $d\Omega^2_{(d-2)}$ is the metric on a unit $(d-2)$-dimensional round sphere. Here
\be\n{2.4a}
F=1-\left(\frac{r_o}{r}\right)^{d-3}\,,
\ee
\be
r_o^{d-3}=\frac{8\Gamma(\frac{d-1}{2})}{(d-2)\pi^{\frac{d-3}{2}}}(M^2+\Sigma^2)^{\tfrac{1}{2}}\,,\n{2.6}
\ee
and
\be
S=\frac{M}{(M^2+\Sigma^2)^{\tfrac{1}{2}}}\,,\n{2.6a}
\ee
where $M\geqslant0$ is the $d$-dimensional Komar mass \cite{MP} measured at asymptotic infinity ($r\to\infty$) and the parameter $\Sigma$ is defined below.

The scalar field, defined up to an additive constant which is irrelevant to our considerations, reads
\be\n{2.8}
\varphi=\frac{\Sigma}{2(M^2+\Sigma^2)^{\tfrac{1}{2}}}\ln|F|\,.
\ee
In the asymptotic region we have
\be\n{2.9}
\varphi\sim-\frac{4\Gamma(\frac{d-1}{2})}{(d-2)\pi^{\frac{d-3}{2}}}\frac{\Sigma}{r^{d-3}}\,.
\ee
Thus, we define $\Sigma\in(-\infty, \infty)$ as the $d$-dimensional ``scalar charge." Hence, expression \eq{2.6a} implies that $S\in [0,1]$ if we take $M\geqslant0$. 

Calculating the energy-momentum tensor components in a local orthonormal frame, we derive the following energy density $\epsilon$ and the principal pressures $p_{\hr}$, $p_{\hal}$: 
\be\n{ppp}
\epsilon=p_{\hr}=-p_{\hal}=\frac{(d-3)\Gamma^2(\frac{d-1}{2})\Sigma^2}{(d-2)\pi^{d-2} r^{2(d-2)}F^{\tfrac{\mathstrut 1-S}{d-3}+1}}\,,
\ee
where the index $\hal=3,...,d$ stands for orthonormal components in the compact dimensions of the $(d-2)$-dimensional round sphere. The scalar field obeys the strong and the dominant energy conditions. Thus, by continuity it obeys the weak and the null energy conditions (see, e.g., \cite{Haw,Visser}).

The Fisher solution has the following limiting cases.

The pure scalar charge case: $M=0$. According to expressions \eq{2.6} and \eq{2.6a}, this case implies
\be
r_o^{d-3}\rvert_{M=0}=r_{\Sigma}^{d-3}=\frac{8\Gamma(\frac{d-1}{2})\Sigma}{(d-2)\pi^{\frac{d-3}{2}}}\,,\hhh S=0. 
\ee
Thus, $\varphi=1/2\ln|F|$, and the corresponding metric is
\ba\n{2.10}
ds^2&=&-dt^2+F_{\Sigma}^{-\tfrac{d-4}{d-3}}dr^2+r^2F_{\Sigma}^{\tfrac{1}{d-3}}d\Omega^2_{(d-2)}\,,
\ea
where 
\be\n{2.10a}
F_{\Sigma}=1-\left(\frac{r_{\Sigma}}{r}\right)^{d-3}\,. 
\ee
We shall call this solution the {\em massless Fisher} solution.
 
The pure mass case: $\Sigma=0$. According to expressions \eq{2.6} and \eq{2.6a}, this case implies 
\be
r_o^{d-3}\rvert_{\Sigma=0}=r_{M}^{d-3}=\frac{8\Gamma(\frac{d-1}{2})M}{(d-2)\pi^{\frac{d-3}{2}}}\,,\hhh S=1. 
\ee
Thus, $\varphi=0$, and the corresponding metric is known as the $d$-dimensional Schwarzschild-Tangherlini black hole \cite{Tan}
\ba\n{2.10b}
ds^2&=&-F_Mdt^2+F_M^{-1}dr^2+r^2d\Omega^2_{(d-2)}\,,
\ea
where 
\be\n{2.10c}
F_M=1-\left(\frac{r_M}{r}\right)^{d-3}\,. 
\ee
The uniqueness of the Schwarzschild-Tangherlini solution was proven in \cite{Gib1,Gib2}.

\subsection{Duality}

The Fisher solution presented above possesses a certain duality symmetry. Here we show that the static, spherically symmetric spacetimes \eq{2.4} corresponding to different values of $M$ and $\Sigma$ are dual to each other. In particular, we show that the Fisher solution is dual to the Schwarzschild-Tangherlini black hole of a particular mass.  

Let us present the metric \eq{2.4} in the following form:
\be\n{met}
ds^2=-k^2dt^2+k^{-\tfrac{2}{d-3}}\bar{g}_{\mu \nu}dx^{\mu}dx^{\nu}.
\ee 
Here, $-k^2$ is the squared norm of the timelike Killing vector $\delta^a_{\,\,\,t}$ and $k^{-2/(d-3)}\bar{g}_{\mu \nu}$ is the $(d-1)$-dimensional spatial metric on a hypersurface orthogonal to $\delta^a_{\,\,\,t}$. We can reduce the $d$-dimensional action \eq{2.1} for the metric \eq{met} to a $(d-1)$-dimensional action for the metric $\bar{g}_{\mu \nu}$. Let us first decompose the Ricci scalar $R$ with respect to a basis defined by the unit timelike vector $k^{-1}\delta^a_{\,\,\,t}$ and $(d-1)$ basis vectors tangential to the hypersurface (see, e.g., \cite{WI}),
\be\n{RI}
R=\widetilde{R}-2\widetilde{\nabla}^{\mu}\widetilde{\nabla}_{\mu}\ln|k|-2k^{\tfrac{2}{d-3}}\bar{g}^{\mu \nu}(ln\,|k|)_{,\mu}\,(ln\,|k|)_{,\nu}.
\ee
Here the $(d-1)$-dimensional Ricci scalar $\widetilde{R}$ and the covariant derivative $\widetilde{\nabla}_{\mu}$ are associated with the metric $k^{-2/(d-3)}\bar{g}_{\mu \nu}$. Applying the conformal transformation defined by the conformal factor $k^{-2/(d-3)}$ to the Ricci scalar $\widetilde{R}$ we derive (see, e.g., \cite{Haw})
\ba\n{RID}
\widetilde{R}&=&k^{\tfrac{2}{d-3}}\left[\bar{R}+\tfrac{2}{d-3}\bar{\nabla}^{\mu}\bar{\nabla}_{\mu}\ln|k|\right.\nonumber\\
&-&\left.\tfrac{d-2}{d-3}\bar{g}^{\mu \nu}(ln\,|k|)_{,\mu}\,(ln\,|k|)_{,\nu}\right].
\ea
Here the $(d-1)$-dimensional Ricci scalar $\bar{R}$ and the covariant derivative $\bar{\nabla}_{\mu}$ are associated with the metric $\bar{g}_{\mu \nu}$. Substituting \eq{RID} into \eq{2.1}, eliminating a surface term, and neglecting an integral over the Killing coordinate $t$ we derive the following $(d-1)$-dimensional action for the metric $\bar{g}_{\mu \nu}$: 
\ba\n{Ac}
\mathcal{S}[\bar{g}_{\mu \nu},\,k,\,\varphi]&=&\frac{1}{16\pi}\int d^{d-1}x\sqrt{\bar{g}}\left(\bar{R}-\tfrac{d-2}{d-3}\bar{g}^{\mu \nu}\right.\nonumber\\
&\times&\Bigl.\bigl[\varphi_{,\mu}\varphi_{,\nu}+(ln\,|k|)_{,\mu}\,(ln\,|k|)_{,\nu}\bigl] \Bigr).
\ea
According to the principle of least action, variation of the action \eq{Ac} with respect to the fields $\bar{g}_{\mu \nu}$, $k$, and $\varphi$ gives the following equations \cite{Woolgar}:
\ba\
&&\bar{R}_{\mu\nu}=\tfrac{d-2}{d-3}\bigl[\varphi_{,\mu}\varphi_{,\nu}+(ln\,|k|)_{,\mu}\,(ln\,|k|)_{,\nu}\bigl],\n{Aca}\\
&&\bar{\nabla}^{\mu}\bar{\nabla}_{\mu}(ln\,|k|)=0,\n{Acb}\\
&&\bar{\nabla}^{\mu}\bar{\nabla}_{\mu}\varphi=k^{-\tfrac{2}{d-3}}\nabla^a\nabla_a\varphi=0.\n{Acc}
\ea
The first equality in Eq. \eq{Acc} holds because the scalar field is static.

We see that the action \eq{Ac} and the field equations \eq{Aca}-\eq{Acc} are invariant under the following transformation:
\be\n{Acd}
\begin{rcase}
ln\,|k'|&=ln\,|k|\cos\psi+\varphi\sin\psi \\
\varphi'&=-ln\,|k|\sin\psi+\varphi\cos\psi\,
\end{rcase},
\ee
which we shall call a {\em duality transformation}. Here the primes denote the dual solution and $\psi$ is the duality transformation parameter whose range is defined below. The duality transformation is analogous to the Buscher T-duality transformation \cite{Ortin}. The metric dual to the metric \eq{met} is
\be\n{Ace}
ds^2=-k'^2dt^2+k'^{-\tfrac{2}{d-3}}\bar{g}_{\mu \nu}dx^{\mu}dx^{\nu}.
\ee   
Thus, we can construct the dual solution \eq{Ace} to the field equations \eq{Aca}-\eq{Acc} if some solution \eq{met} is already known. In particular, we can apply the duality transformations \eq{Acd} to generate the Fisher solution \eq{2.4} without integration of the Einstein equations, starting from the Schwarzschild-Tangherlini metric \eq{2.10b} with $r_{M}=r_o$ and taking $\cos\psi=S$. This procedure suggests that we can present the duality transformation \eq{Acd} in different form, in terms of the mass $M$ and the scalar charge $\Sigma$. Indeed, starting from the metric \eq{2.4} we have $k^2=F^{S}$. Using expressions \eq{2.4a}-\eq{2.8} and \eq{Acd} we find that $r_o'=r_o$. Thus, $r_o$ [see, \eq{2.6}] is invariant of the duality transformation (there are other invariants of the duality transformation which we present in Sec. VI). Hence, we can present the duality transformation \eq{Acd} in the following form:
\be\n{2.11}
\begin{rcase}
M'&=M\cos\psi+\Sigma\sin\psi \\
\Sigma'&=-M\sin\psi+\Sigma\cos\psi\,
\end{rcase}.
\ee
Thus, we have the duality transformation between the mass and the scalar charge acting in the parameter space $(M,\Sigma)$. To define the range for $\psi$ we consider dual Fisher solutions which have nonnegative mass $M\geqslant0$. Thus, for a Fisher solution defined by the parameters $(M_o\geqslant0,\Sigma_o)$ such that
\be
\psi_o=\arctan(\Sigma_o/M_o)\in[-\pi/2,\pi/2]\,,
\ee
the corresponding duality transformation parameter is defined by 
\be
\psi\in[-\pi/2+\psi_o, \pi/2+\psi_o]\,.
\ee
In particular, for $\psi=\mp\pi/2+\psi_o$ we have $M'_o=0$ and $\Sigma'_o=\pm(M_o^2+\Sigma_o^2)^{1/2}$, which is a massless Fisher solution \eq{2.10} with $r_{\Sigma}=r_o$. For $\psi=\psi_o$ we have $M'_o=(M_o^2+\Sigma_o^2)^{1/2}$ and $\Sigma'_o=0$, which is a Schwarzschild-Tangherlini black hole \eq{2.10b} with $r_M=r_o$. Here and in what follows, unless stated explicitly, we shall refer to the massless Fisher solution \eq{2.10} and to the Schwarzschild-Tangherlini black hole \eq{2.10b} having in mind their {\em dual to the Fisher solution form}, i.e., for $r_{\Sigma}=r_o$ and for $r_M=r_o$, respectively. This convention can be expressed in the following way:
\ba
r_{\Sigma}&=&r_o\Longleftrightarrow \Sigma'=(M^2+\Sigma^2)^{\tfrac{1}{2}}\,,\hhh M'=0\,,\n{mf}\\
r_{M}&=&r_o\Longleftrightarrow M'=(M^2+\Sigma^2)^{\tfrac{1}{2}}\,,\hhh \Sigma'=0\,.\n{st}
\ea

\begin{figure}[htb]
\begin{center}
\includegraphics[height=5.41cm,width=7cm]{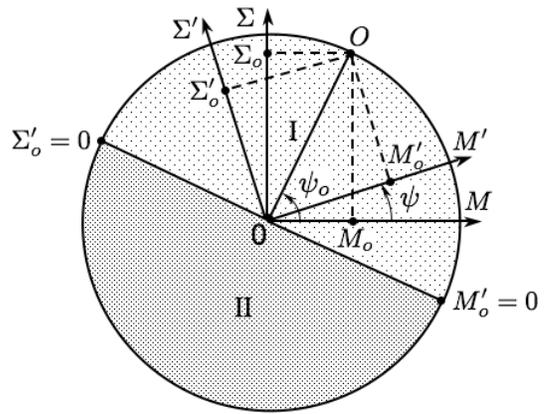}
\caption{Duality diagram. Point $O$ represents the Fisher solution defined by the mass $M_o$ and the scalar charge $\Sigma_o$. Sector I represents its dual nonnegative mass solutions ($M'_o\geqslant0$). One such dual Fisher solution is defined by the mass $M'_o$ and the scalar charge $\Sigma'_o$.  Sector II represents dual negative mass solutions ($M'_o<0$) which we do not consider here.}\label{D1}
\end{center}
\end{figure}

The duality transformation \eq{2.11} is illustrated in Fig.~\ref{D1}. From the duality diagram we see that increase (decrease) in the mass $M'$ corresponds to decrease (increase) in the scalar charge $\Sigma'$. Thus, the duality transformation can be considered as a change of the mass $M$ and the scalar charge $\Sigma$ in the original solution to their dual values $M'$ and $\Sigma'$. From this point of view, the duality transformation is a mapping between different members of the Fisher family of solutions $(M,\Sigma)$. In particular, for $\psi_o=0$, and $\psi=\pi/2$ the Schwarzschild-Tangherlini black hole and the massless Fisher solution are dual to each other (see, \cite{Ortin}, p. 216). In general, any Fisher solution is dual to the Schwarzschild-Tangherlini black hole.  

The duality transformation \eq{2.11} is a transformation between different solutions of the same theory \eq{2.1}. Each of these solutions represents a spacetime of certain properties. That is, all these spacetimes are spherically symmetric, static, and asymptotically flat. However, there is an essential difference between the Schwarzschild-Tangherlini spacetime and the Fisher solution. The Schwarzschild-Tangherlini spacetime represents a black hole of the mass $M'$ whose event horizon is defined by $r=r_o$. The horizon is regular and the spacetime singularity is located behind the horizon at $r=0$. However, as we shall see in the next section, the Fisher spacetime does not have an event horizon but instead has a naked singularity located at $r=r_o$. In what follows, we shall study the properties of the Fisher solution. We shall see that the spacetime geometry near the naked singularity has interesting properties which may be seen as a manifestation of the duality.  

\subsection{The Fisher universe}   

As we already mentioned, $r=r_o$ is a naked curvature singularity of the Fisher solution. Thus, we have to cut $r=r_o$ out of the Fisher manifold defined by the coordinates $(t,r,x^{\alpha})$, where the index $\alpha=3,...,d$ stands for compact coordinates which define the position of a point on a unit $(d-2)$-dimensional round sphere. As we shall see, $r=0$ is another curvature singularity of the Fisher solution. Thus, the cut divides the Fisher manifold into two disconnected parts defined by $r\in(r_o,\infty)$ and $r\in(0,r_o)$. In what follows, we shall call the region $r\in(r_o,\infty)$ the {\em Fisher spacetime}, and the region $r\in(0,r_o)$ the {\em Fisher universe}. 

In a traditional approach, one considers that part of a manifold which represents the external field due to some source and which is asymptotically flat, if such exists. Such an approach was taken before in the case of the Fisher solution (see, e.g., \cite{Wym,Agn,Rob}). Here we shall consider both the parts of the manifold. The reason for such a consideration is motivated by the duality between the Schwarzschild-Tangherlini black hole and the Fisher solution which we discussed above. In particular, the interior of the Schwarzschild-Tangherlini black hole corresponds to $r\in(0,r_o)$. Thus, to consider a dual to the interior part we have to consider the region $r\in(0,r_o)$ of the Fisher solution. However, for $r\in(0,r_o)$ and nonzero scalar charge the metric \eq{2.4} is in general complex valued due to noninteger exponents \cite{complex}. One can make the metric real valued by introducing absolute values $|F|$ into the metric functions in an appropriate way. Such a modified metric solves the Einstein equations \eq{A2a}-\eq{A2c} but has the signature $-(d-2)$. As a result, for $r\in(0,r_o)$ the periodic angular coordinate becomes timelike which leads to causality violation, which we would not like to have here. There is another way to make the metric real valued in the region, which is to replace $r_o^{d-3}$ with $r_o^{d-3}\,\text{sign}(r-r_o)$ in the metric functions. However, such a choice implies that the dual Schwarzschild-Tangherlini black hole has negative mass $M'<0$, which is out of our consideration. However, there is yet another way to get a real valued metric for $r\in(0,r_o)$.  Namely, one can apply complex transformations preserving the signature of the metric and keeping mass nonnegative. The following complex transformations bring the metric in the region $r\in(0,r_o)$ to a real valued form:
\be\n{3.1}
\begin{rcase}
t&=(-1)^{\tfrac{\mathstrut 1-S}{2}}\tau\\
r&=(-1)^{\tfrac{\mathstrut S-1}{2(d-3)}}\rho\\
\end{rcase},\hhh
\begin{rcase}
M&=(-1)^{\tfrac{\mathstrut S-1}{2}}\mu\\
\Sigma&=(-1)^{\tfrac{\mathstrut S-1}{2}}\sigma\,
\end{rcase}.
\ee
Note that $r_o$ transforms like $r$ and according to expression \eq{2.6a}, $S$ is an invariant,
\be\n{3.1a}
S=\frac{M}{(M^2+\Sigma^2)^{\tfrac{1}{2}}}=\frac{\mu}{(\mu^2+\sigma^2)^{\tfrac{1}{2}}}\,.
\ee
In the limit $S\to1$ these transformations become merely a relabeling of the coordinates and parameters and preserve the positive direction of the time and space coordinates. In addition, in the limit $S\to1$ the two disconnected parts of the Fisher manifold represent the exterior and interior of the Schwarzschild-Tangherlini black hole, and can be analytically extended to a larger manifold which represents the maximal $d$-dimensional extension of the Schwarzschild-Tangherlini solution. Such an extension was given in the Kruskal coordinates in \cite{GL} and in another coordinate system in \cite{Lake1,Lake2}.  
  
Applying the transformations \eq{3.1} to the metric \eq{2.4} we derive
\ba\n{3.2}
ds^2=\Phi^{S}d\tau^2-\Phi^{\tfrac{\mathstrut 1-S}{d-3}-1}d\rho^2+\rho^2\Phi^{\tfrac{\mathstrut1-S}{d-3}}d\Omega^2_{(d-2)},
\ea
where
\be\n{3.2a}
\Phi=\left(\frac{\rho_o}{\rho}\right)^{d-3}-1\,,\hhh\rho_o^{d-3}=\frac{8\Gamma(\frac{d-1}{2})}{(d-2)\pi^{\frac{d-3}{2}}}(\mu^2+\sigma^2)^{\tfrac{1}{2}}\,.
\ee
Here the compact coordinate $\rho\in(0,\rho_o)$ is timelike. The spacetime \eq{3.2} represents an anisotropic universe which we call the Fisher universe. We shall study properties of the Fisher universe in the following sections. 

Applying the transformations \eq{3.1} to the scalar field \eq{2.8} we derive
\be\n{3.2b}
\varphi=\frac{\sigma}{2(\mu^2+\sigma^2)^{\tfrac{1}{2}}}\ln\Phi\,.
\ee
Calculating the energy-momentum tensor components of the scalar field in a local orthonormal frame we derive the following energy density $\epsilon$ and the principal pressures $p_{\hta}$, $p_{\hal}$ [cf. Eq.\eq{ppp}]: 
\be\n{3.2c}
p_{\hta}=\epsilon=p_{\hal}=\frac{(d-3)\Gamma^2(\frac{d-1}{2})\sigma^2}{(d-2)\pi^{d-2}\rho^{2(d-2)}\Phi^{\tfrac{\mathstrut 1-S}{d-3}+1}}\,.
\ee
Thus, the scalar field represents a stiff fluid. It obeys the strong and the dominant energy conditions. Therefore, by continuity it obeys the weak and the null energy conditions.

In the case of the massless Fisher solution \eq{2.10}, the transformation of the $t$ coordinate in \eq{3.1} is the Wick rotation. This case implies
\be\n{3.2d}
\rho_o^{d-3}\rvert_{\mu=0}=\rho_{\sigma}^{d-3}=\frac{8\Gamma(\frac{d-1}{2})\sigma}{(d-2)\pi^{\frac{d-3}{2}}}\,,\hhh S=0. 
\ee
Thus, $\varphi=1/2\ln\Phi$, and the corresponding metric is
\ba\n{3.2e}
ds^2&=&d\tau^2-\Phi_{\sigma}^{-\tfrac{d-4}{d-3}}d\rho^2+\rho^2\Phi_{\sigma}^{\tfrac{1}{d-3}}d\Omega^2_{(d-2)}\,,
\ea
where 
\be\n{3.2f}
\Phi_{\sigma}=\left(\frac{\rho_{\sigma}}{\rho}\right)^{d-3}-1\,. 
\ee
We shall call this solution the {\em massless Fisher universe}. Analogous to \eq{mf} the dual to the Fisher universe massless solution corresponds to
\be\n{mfu}
\rho_{\sigma}=\rho_o\Longleftrightarrow \sigma'=(\mu^2+\sigma^2)^{\tfrac{1}{2}}\,,\hhh \mu'=0\,.
\ee
Here and in what follows, unless stated explicitly, we shall refer to the massless Fisher universe \eq{3.2e} having in mind the dual to the Fisher universe form \eq{mfu}.

In general, the mass transformation in \eq{3.1} for arbitrary $S\in[0,1]$ has the following form:
\be\n{3.3}
\mu=M\cos\left(\frac{\pi}{2}[1-S]\right)+iM\sin\left(\frac{\pi}{2}[1-S]\right)\,,
\ee
where the first term is the bradyon mass $M_B$ and the second term is the tachyon mass $M_T$. In these notations, $S$ defines the ratio of the tachyon mass to the bradyon mass as follows:
\be\n{3.6}
\frac{M_T}{M_B}=i\tan\left(\frac{\pi}{2}[1-S]\right)\,.
\ee
The scalar charge transformation in \eq{3.1} is analogical to \eq{3.3},
\be\n{3.3a}
\sigma=\Sigma\cos\left(\frac{\pi}{2}[1-S]\right)+i\Sigma\sin\left(\frac{\pi}{2}[1-S]\right)\,,
\ee
where the first term is a real scalar field charge and the second term is a ghost scalar field charge. However, expressions \eq{3.3} and \eq{3.3a} are merely transformations. It is not clear if they have any physical meaning. In the Fisher spacetime and the Fisher universe the mass and the scalar charge are real. 

\section{Curvature singularities}

\subsection{Spacetime invariants}

Spacetime curvature singularities, like those located inside of black holes, are associated with infinitely growing spacetime curvature invariants. To determine singularities of the Fisher solution we calculate the Ricci scalar and the Kretschmann invariant. The Ricci scalar is
\be\n{3.7}
R=\frac{1-S^2}{4}\frac{r_o^{2(d-3)}}{r^{S+d-2}}\frac{(d-2)(d-3)}{\left(r^{d-3}-r_o^{d-3}\right)^{\tfrac{\mathstrut 1-S}{d-3}+1}}\,.
\ee
We see that the Ricci scalar diverges at $r=r_o$, if $S\ne1$, and at $r=0$. According to the transformations \eq{3.1}, $r=r_o$ and $r=0$ correspond to $\rho=\rho_o$ and $\rho=0$, respectively. The Schwarzschild-Tangherlini black hole solution ($S=1$) is Ricci flat.

For $S\ne1$ the Kretschmann invariant presented in Appendix B is proportional to $R^2$, therefore, it diverges at the same points. For the Schwarzschild-Tangherlini black hole the Kretschmann scalar is
\be\n{3.9}
\mathcal{K}=\frac{r_o^{2(d-3)}}{r^{2(d-1)}}(d-1)(d-2)^2(d-3)\,.
\ee
It diverges at $r=0$. The analysis of the spacetime invariants shows that the Fisher solution is singular at $r=r_o$ ($\rho=\rho_o$) for $S\in[0,1)$ and at $r=\rho=0$ for $S=[0,1]$. Both the singularities are {\em central}, i.e., the corresponding areal radii vanish at the singularities [see, expressions \eq{4.6b} and \eq{4.8b}]. We shall study the properties of these singularities.

\subsection{Strength of the singularities}

Spacetime curvature singularities can be characterized according to their strength. A definition of singularity strength based on purely geometric properties of spacetime was proposed in \cite{Tip}. According to that definition, there are two types of curvature singularities, gravitationally {\em weak} and {\em strong}. Namely, if a volume (an area) element defined by linearly independent spacelike vorticity-free Jacobi fields propagating along any timelike (null) geodesic and orthogonal to its tangent vector vanishes at spacetime singularity, the singularity is called strong, otherwise, if the volume (the area) element does not vanish and remains finite, the singularity is called weak. Necessary and sufficient conditions for strong curvature singularities were formulated in \cite{CK,CJS}. The definition above was subsequently modified in \cite{Nolan}, where behavior of each Jacobi field was taken into account. According to the renewed definition, a spacetime singularity is called strong if at least one Jacobi field vanishes or diverges at the singularity. For example, a singularity is called strong if some of the Jacobi fields diverge and others vanish such that the volume element remains finite at the singularity. A {\em deformationally strong} singularity was defined in \cite{Ori}. According to that definition, a spacetime singularity is called deformationally strong if the volume element diverges, or at least one Jacobi field diverges, but the volume element remains finite, for other Jacobi fields vanish at the singularity.

Here we shall study the strength of the Fisher spacetime and the Fisher universe singularities. Let us begin with the Fisher spacetime \eq{2.4}, $r\in(r_o,\infty)$. We shall study behavior of Jacobi fields defined for radial timelike and null geodesics near the spacetime singularity located at $r=r_o$. Equations for the geodesic motion can be derived from the corresponding Lagrangian $\mathcal{L}$ associated with the metric \eq{2.4}, 
 \ba\n{3.11}
2\mathcal{L}&=&-F^{S}\dot{t}^2+F^{\tfrac{\mathstrut 1-S}{d-3}-1}\dot{r}^2=\varepsilon\,,
\ea
where $\varepsilon$ is equal to $-1$ for timelike and $0$ for null geodesics. The overdot denotes the differentiation with respect to $\lambda$ which is the proper time for timelike and the affine parameter for null geodesics. We define $\lambda$ such that the geodesics approach the singularity located at $r=r_o$ as $\lambda\to-0$. The radial geodesics are defined by the unit tangent vector $k^{a}=\dot{x}^a$ whose nonzero components in a local orthonormal frame are given by
\ba
k^{\htt}&=&F^{\tfrac{\mathstrut S}{2}}\,\dot{t}=E_\varepsilon\,F^{-\tfrac{\mathstrut S}{2}}\,,\n{3.12a}\\
k^{\hr}&=&F^{\tfrac{\mathstrut 1-S}{2(d-3)}-\tfrac{1}{2}}\,\dot{r}=\pm\left[(k^{\htt})^2+\varepsilon \right]^{\tfrac{1}{2}}\,,\n{3.12b}
\ea 
where ``$+$" stands for outgoing and ``$-$" stands for ingoing geodesics, and $E_\varepsilon=const$ which we define as follows:
\be
E_{-1}>1,\hhh E_0=1\,.
\ee
We consider ingoing geodesics. One can check that the radial geodesics approach the singularity for finite values of $\lambda$. For $S\in(0,1]$ the geodesics approach $r=r_o$ in infinite coordinate time $t$ which measures proper time of an observer which is at rest with respect to the gravitational center (the naked singularity) and located at asymptotic infinity $(r\to\infty)$. For $S=0$ the coordinate time $t$ is finite.

Jacobi fields $Z^{\ha}(\lambda)$ are orthogonal to $k^{\ha}$ and represent the spatial separation of two points of equal values of $\lambda$ located on neighboring geodesics. They satisfy the Jacobi geodesic deviation equation (see, e.g., \cite{Haw})
\be\n{3.13}
\ddot{Z}^{\ha}+R_{\hc\hb\hd}^{\ \ \ \ \ \ha}Z^{\hb}k^{\hc} k^{\hd}=0\,,
\ee 
where $R_{\hc\hb\hd}^{\ \ \ \ \ha}$ are the Riemann tensor components defined in the local orthonormal frame (see, Appendix B). 

For radial timelike geodesics we define two types of the Jacobi fields. The {\em radial} Jacobi field
\be\n{rad}
Z^{\eta}\partial_{\eta}=Z^{\htt}\partial_{\htt}+Z^{\hr}\partial_{\hr}\,,\hhh g_{\eta\eta}=1\,,
\ee
and the $(d-2)$ orthogonal {\em angular} Jacobi fields 
\be\n{ang}
Z^{\hal}\partial_{\hal}\,,\hhh g_{\hal\hal}=1\,,\hhh \hal=3,...,d\,.
\ee
The spacelike vectors $\partial_{\eta},\partial_{\hal},\,\hal=3,...,d\,$ form a $(d-1)$-dimensional orthonormal basis which is parallel propagated along the radial timelike geodesics. As far as we are interested in spatial separations of neighboring geodesics, for radial null geodesics we consider only the angular Jacobi fields \eq{ang}.  

The radial Jacobi field satisfies the Jacobi equation
\be\n{3.14}
\ddot{Z}^{\eta}+R_{\htt\hr\htt}^{\ \ \ \ \hr}Z^{\eta}=0\,.
\ee
Approximating expressions \eq{3.12a}, \eq{3.12b}, and \eq{B1a} near the singularity we derive
\be\n{3.15}
\ddot{Z}^{\eta}-C\vert \lambda\vert^{-2+\tfrac{\mathstrut 2S(d-3)}{d-2+S(d-4)}}Z^{\eta} \approx 0\,,
\ee
where
\be\n{3.16}
C=\frac{S(1-S)(d-2)(d-3)(2r_o)^{-\tfrac{\mathstrut 2S(d-3)}{d-2+S(d-4)}}}{\left(E_{-1}[d-2+S(d-4)]\right)^{2-\tfrac{\mathstrut 2S(d-3)}{d-2+S(d-4)}}}\,.
\ee
This equation is a particular case of the Emden-Fowler equation (see, Eq. (2.1.2.7), p.132 in \cite{Pol}). Its solutions are  expressed in terms of the modified Bessel functions of the first and second kind. Using asymptotics of the modified Bessel functions for small values of their arguments (see, e.g., Eqs. (9.6.7) and (9.6.9) in \cite{AS}) we derive the asymptotic behavior of the radial Jacobi field near the singularity
\be\n{3.17}
Z^{\eta}(\lambda)\sim c_1+c_2\vert \lambda \vert\sim c_1\,.
\ee
Here and in what follows $c_{1,2}=const\ne0$. Thus, for $S\in[0,1)$ the radial Jacobi field remains finite at the singularity. Although it is obvious that the Jacobi field is finite in the case of the Schwarzschild-Tangherlini black hole $(S=1)$, for there is no spacetime singularity at $r=r_o$, it is remarkable that the radial Jacobi field is finite at the singularity of the Fisher spacetime. Thus, the singularity at $r=r_o$ is of a special type, which we call {\em radially weak}.    

Now we consider the angular Jacobi fields \eq{ang}. Each of the $(d-2)$ angular Jacobi fields $Z^{\hal}$ satisfies the following equation (no summation over $\hal$):
\be\n{3.18}
\ddot{Z}^{\hal}+\left[R_{\htt\hal\htt}^{\ \ \ \ \hal}(k^{\htt})^2+R_{\hr\hal\hr}^{\ \ \ \ \hal}(k^{\hr})^2\right]{Z}^{\hal}=0\,.
\ee
This equation is valid for both the radial timelike and null geodesics. Approximating expressions \eq{3.12a}, \eq{3.12b}, \eq{B1b}, and \eq{B1c} near the singularity and applying the method of Frobenius we derive the asymptotic behavior of the angular Jacobi fields
\ba\n{3.19}
Z^{\hal}(\lambda)&\sim& c_1\vert \lambda \vert^{\tfrac{\mathstrut 1-S}{d-2+S(d-4)}}+c_2\vert \lambda \vert^{\tfrac{\mathstrut (1+S)(d-3)}{d-2+S(d-4)}}\nonumber\\
&\sim& c_1\vert \lambda \vert^{\tfrac{\mathstrut 1-S}{d-2+S(d-4)}}\,.
\ea 
This expression is valid for the radial timelike and null geodesics for $S\in[0,1]$. There is no singularity for $S=1$, and the corresponding angular Jacobi fields are finite. For other values of $S$ the angular Jacobi fields vanish. 

Let us now study the singularities of the Fisher universe \eq{3.2}. We shall study behavior of the Jacobi fields defined for radial timelike and null geodesics approaching the spacetime singularities located at $\rho=\rho_o$ and at $\rho=0$. Applying the transformations \eq{3.1} to expressions \eq{3.12a} and \eq{3.12b} we derive the nonzero components of the unit tangent vector
\ba
k^{\hta}&=&\Phi^{\tfrac{\mathstrut S}{2}}\,\dot{\tau}=\mathscr{E}_\varepsilon\,\Phi^{-\tfrac{\mathstrut S}{2}}\,,\n{3.19a}\\
k^{\hrho}&=&\Phi^{\tfrac{\mathstrut 1-S}{2(d-3)}-\tfrac{1}{2}}\,\dot{\rho}=\mp\left[(k^{\hta})^2-\varepsilon \right]^{\tfrac{1}{2}}\,,\n{3.19b}
\ea
where ``$-$" stands for outgoing and ``$+$" stands for ingoing geodesics and $\mathscr{E}_\varepsilon=const$ which we define as follows:
\be
\mathscr{E}_{-1}\geqslant0,\hhh \mathscr{E}_0=1\,.
\ee
One can check that the radial geodesics approach the singularities for finite values of $\lambda$. For $S\in(0,1]$ and geodesics approaching $\rho=\rho_o$, the finite change of $\lambda$ corresponds to an infinite change of the spacelike coordinate $\tau$ for $\mathscr{E}_{-1}>0$, whereas for geodesics approaching $\rho=0$ the change of the spacelike coordinate $\tau$ vanishes. For $S=0$ the change of the coordinate $\tau$ is always finite.

The geodesics deviation equations for the radial and angular Jacobi fields \eq{rad} and \eq{ang} orthogonal to the tangent vector \eq{3.19a} and \eq{3.19b} can be constructed by applying the transformations \eq{3.1} to the Riemann tensor components in Eqs. \eq{3.14} and \eq{3.18}.  Solving the derived equations near the singularity $\rho=\rho_o$ of the Fisher universe, one can see that the behavior of the Jacobi fields is exactly the same as the behavior of the corresponding Jacobi fields \eq{3.17} and \eq{3.19} near the singularity $r=r_o$ of the Fisher spacetime. 

Let us examine the singularity at $\rho=0$. Approximating the Jacobi equation \eq{3.14} near the singularity and applying the method of Frobenius, we derive the asymptotic behavior of the radial Jacobi field,
\ba\n{3.20}
Z^{\eta}(\lambda)&\sim& c_1\vert \lambda \vert^{-\tfrac{\mathstrut S(d-3)}{d-2+S}}+c_2\vert \lambda \vert^{\tfrac{\mathstrut (1+S)(d-2)}{d-2+S}}\nonumber\\
&\sim& c_1\vert \lambda \vert^{-\tfrac{\mathstrut S(d-3)}{d-2+S}}\,,
\ea
where $S\in(0,1]$. Thus, as in the case of the Schwarzschild-Tangherlini black hole, the radial Jacobi field diverges. However, in the case of the massless Fisher solution ($S=0$) the radial Jacobi field is finite at the singularity and given by expression \eq{3.17}. Thus, this singularity is radially weak as well.

Let us consider the asymptotic behavior of the angular Jacobi fields \eq{ang} corresponding to the radial timelike and null geodesics approaching the singularity. For timelike geodesics and for $d=4$ we have
\be\n{3.24}
Z^{\hal}(\lambda)\sim c_1\vert \lambda \vert^{\tfrac{1}{2+S}}+c_2\vert \lambda \vert^{\tfrac{\mathstrut 1+S}{2+S}}\sim c_1\vert \lambda \vert^{\tfrac{1}{2+S}}\,,
\ee  
whereas for $d>4$ we have
\ba\n{3.25}
Z^{\hal}(\lambda)&\sim& c_1\vert \lambda \vert^{\tfrac{\mathstrut 1+S}{d-2+S}}+c_2\vert \lambda \vert^{\tfrac{d-3}{d-2+S}}\nonumber\\
&\sim& c_1\vert \lambda \vert^{\tfrac{\mathstrut 1+S}{d-2+S}}\,.
\ea
For null geodesics we have 
\ba\n{3.22}
Z^{\hal}(\lambda)&\sim& c_1\vert \lambda \vert^{\tfrac{\mathstrut 1+S}{d-2-S(d-4)}}+c_2\vert \lambda \vert^{\tfrac{\mathstrut (1-S)(d-3)}{d-2-S(d-4)}}\nonumber\\
&\sim& c_1\vert \lambda \vert^{\tfrac{\mathstrut 1+S}{d-2-S(d-4)}},\hhh S\in\left(0,\tfrac{d-4}{d-2}\right],
\ea 
and
\be\n{3.23}
Z^{\hal}(\lambda)\sim c_2\vert \lambda \vert^{\tfrac{\mathstrut (1-S)(d-3)}{d-2-S(d-4)}},\hhh S\in\left(\tfrac{d-4}{d-2},1\right].
\ee  
Thus, for $S\in (0,1]$ and the radial timelike and null geodesics approaching the singularity at $\rho=0$, the angular Jacobi fields vanish.

To define the strength of the singularities we calculate first the norm of the $(d-1)$-dimensional volume element of a synchronous frame which is defined by 1-forms corresponding to the radial and angular Jacobi fields calculated for the radial timelike geodesics as follows: 
\be\n{3.26}
\lVert V_{(d-1)}\rVert=\lvert Z^{\eta}\rvert\prod_{\hal=3}^{d}\lvert Z^{\hal}\rvert\,. 
\ee
Near the singularities the norm of the volume element can be approximated according to the behavior of the Jacobi fields [see Eqs. \eq{3.17},\eq{3.19}, and \eq{3.20}-\eq{3.25}] as follows:
\be
\lVert V_{(d-1)}\lVert\sim|\lambda|^v\,,
\ee
where the exponent $v=const$ defines how fast the norm of the volume element vanishes or diverges when we approach the singularities $(\lambda\to-0)$. Thus, to compare the strength of the singularities of the Fisher spacetime and the Fisher universe we compare the corresponding values of the exponent $v$. The results are given in Table I.
\begin{table}
\caption{\label{tab:table1} The values of the exponent $v$ for the radial timelike geodesics approaching the singularities.}
\begin{ruledtabular}
\begin{tabular}{ccc}
$d$&$\,\,\,r=r_o\,(\rho=\rho_o)$&$\rho=0$\\
\hline
$=$\,4&$1-S\geqslant0$&$\frac{\mathstrut 2-S}{2+S}\geqslant\frac{1}{3}$\\
$>$\,4&$\frac{(1-S)(d-2)}{d-2+S(d-4)}\geqslant0$&1\\
\end{tabular}
\end{ruledtabular}
\end{table}
\begin{table}
\caption{\label{tab:table2} The values of the exponent $a$ for the radial null geodesics approaching the singularities.}
\begin{ruledtabular}
\begin{tabular}{cccc}
$d$&$\,\,\,r=r_o\,(\rho=\rho_o)$&$\rho=0$\footnote{Here $S\in\left[0,\frac{d-4}{d-2}\right]$. $^b$Here $S\in\left(\frac{d-4}{d-2},1\right]$.}&$\rho=0^b$\\
\hline
$=$\,4&$1-S\geqslant0$&1&$1-S\geqslant0$\\
$>$\,4&$\frac{\mathstrut (1-S)(d-2)}{d-2+S(d-4)}\geqslant0$&$\frac{\mathstrut (1+S)(d-2)}{d-2-S(d-4)}\geqslant1$&$\frac{\mathstrut (1-S)(d-2)(d-3)}{d-2-S(d-4)}\geqslant0$\\
\end{tabular}
\end{ruledtabular}
\end{table}
 
For null geodesics approaching the singularities we calculate the norm of the $(d-2)$-dimensional area element which is defined by 1-forms corresponding to the angular Jacobi fields calculated for the radial null geodesics as follows: 
\be\n{3.27}
\lVert A_{(d-2)}\rVert=\prod_{\hal=3}^{d}\lvert Z^{\hal}\rvert. 
\ee
Analogous to the norm of the volume element, the norm of the area element can be approximated near the singularities according to the behavior of the angular Jacobi fields [see Eqs. \eq{3.19},\eq{3.22}, and \eq{3.23}] as follows:
\be
\lVert A_{(d-1)}\rVert\sim|\lambda|^a\,,
\ee
where the exponent $a=const$ defines how fast the norm of the area element vanishes or diverges when we approach the singularities $(\lambda\to-0)$. The values of the exponent $a$ calculated for the radial null geodesics approaching the singularities of the Fisher spacetime and the Fisher universe are given in Table II.

Now we can summarize our results. According to the values of the exponents $v$ and $a$ presented in Tables I and II the volume and the area elements vanish at the singularities, except for the case of $S=1$ and $r=r_o$, where $v=a=0$, so the volume element is finite. This case corresponds to the event horizon of the Schwarzschild-Tangherlini black hole. At the black hole singularity ($r=\rho=0$) the area element is finite as well $(a=0)$. Thus, according to the classifications of spacetime singularities, the singularities of the Fisher spacetime and the Fisher universe are strong. In addition, the strength of the singularity at $\rho=0$ is greater if the value of $S$ is smaller. However, for the radial timelike geodesics and $d>4$ the strength does not depend on $S$. Thus, in general, the scalar field decreases the values of the volume and the area elements. From the tables we see that for $S\in(0,1)$ the singularity at $\rho=0$ is stronger than the singularity at $r=r_o$, whereas for $S=0$ these singularities have equal strength. 

Let us analyze the behavior of the Jacobi fields. An analysis of the angular Jacobi fields \eq{3.19},\eq{3.24},\eq{3.22}, and \eq{3.23} shows that the scalar field contracts the spacetime in the angular directions. However, for the radial timelike geodesics and $d>4$ [see, \eq{3.25}] it decreases the spacetime contraction in the angular directions caused by the gravitational field. From expressions \eq{3.24} and \eq{3.25} we see that in the case of the Schwarzschild-Tangherlini black hole $(S=1)$ the angular Jacobi fields contract faster for $d=5,6$ than for $d=4$, and for $d=4$ and $d=7$ the contraction rates are the same, whereas for $d>7$ the contraction is less than for $d=4$. In the presence of the scalar field $(S\ne1)$ for $d>4$ the contraction is less [see, \eq{3.25}]. An analysis of the radial Jacobi field \eq{3.20} shows that the scalar field decreases its divergency, i.e., the scalar field contracts the Fisher spacetime in the radial direction as well. However, the radial Jacobi fields \eq{3.17} at the singularities at $r=r_o$ and at $\rho=\rho_o$ for $S\in[0,1)$, as well as at the singularity at $\rho=0$ for $S=0$ remain finite. According to our calculations, this is a generic property of the singularities which is valid for any set of initial data. In other words, no fine-tuning is required for such a behavior of the radial Jacobi fields. It implies that a 1-dimensional object, for example, an infinitesimally thin rod, which is moving along a radial timelike geodesic will arrive intact to the singularities without being contracted to zero or stretched to infinity. We call these singularities radially weak. 
 
Finite, nonzero values of the radial Jacobi fields terminating at the radially weak singularities may suggest a $C^0$ local extension \cite{Tip} of the 2-dimensional $(t,r)$ and $(\tau,\rho)$ spacetime surfaces through the singularities. In Sec. VII we shall discuss such an extension for the singularities of the Fisher solution.       

\section{Causal Properties of the Fisher solution}

\subsection{Closed trapped surfaces}

The concept of a {\em closed trapped surface} introduced by Penrose \cite{Pen} was crucial for the formulation of the singularity theorems \cite{Haw}. In a $d$-dimensional spacetime a closed trapped surface $\mathcal{T}$ is a $(d-2)$-dimensional spacelike compact surface without boundary which is defined according to the following property: future directed outgoing and ingoing null geodesics orthogonal to $\mathcal{T}$ are converging at  $\mathcal{T}$. Mathematically, this property is expressed in the following way. Let $\vec{n}^{\pm}$ be future directed null vectors orthogonal to $\mathcal{T}$ and normalized in the following way: $g(\vec{n}^+,\vec{n}^-)=-1$, where ``$+$" stands for outgoing and ``$-$" stands for ingoing null geodesics. Then, the scale-invariant {\em trapping scalar} defined on $\mathcal{T}$ is as follows:
\be\n{4.1}
\Theta_{ \mathcal{T}}=\theta^+\theta^-
\ee
must be positive (see, e.g., \cite{Haw,Sen}). Here $\theta^{\pm}$ are the null expansions of the null geodesics defined on $\mathcal{T}$ and expressed in terms of the null second fundamental forms 
\be
\chi^{\pm}_{\alpha\beta}=e_{(\alpha)}^{\,\,\,a}e_{(\beta)}^{\,\,\,b}\nabla_bn^{\pm}_a,
\ee 
in the following way:
\be\n{4.2}
\theta^{\pm}=\left.\gamma^{\alpha\beta}\chi^{\pm}_{\alpha\beta}\right\vert_{\mathcal{T}}.
\ee    
Here $e_{(\alpha)}^{\,\,\,a},\,\alpha=3,...,d\,$ are the base-vectors tangential to $\mathcal{T}$ and
\be 
\gamma_{\alpha\beta}=\left.g_{ab}\right\vert_{\mathcal{T}}e_{(\alpha)}^{\,\,\,a}e_{(\beta)}^{\,\,\,b}
\ee
is the positive-defined metric induced on $\mathcal{T}$.

Let us examine if closed trapped surfaces are present in the Fisher spacetime and/or the Fisher universe. The Fisher spacetime \eq{2.4}, $r\in(r_o,\infty)$ is static and spherically symmetric. Thus, we define $\mathcal{T}$ by $t=const$ and $r=const$. In this case, the trapping scalar \eq{4.1} is 
\be\n{4.3}
\Theta_{ \mathcal{T}}=-\left.\frac{g^{rr}}{8}\left(\frac{\partial\ln\gamma}{\partial r}\right)^2\right\vert_{r=const}\,,
\ee
where $\gamma=det(\gamma_{\alpha\beta})$ and the indices $\alpha,\beta=3,...,d\,$ stand for angular coordinates. For the Fisher spacetime \eq{2.4}, $r\in(r_o,\infty)$ expression \eq{4.3} reads
\be\n{4.3a}
\Theta_{ \mathcal{T}}=-\left.\frac{(d-2)^2}{8r^{S+d-2}}\frac{\left(2r^{d-3}-(1+S)r_o^{d-3}\right)^2}{\left(r^{d-3}-r_o^{d-3}\right)^{\tfrac{\mathstrut 1-S}{d-3}+1}}\right\vert_{r=const.}
\ee
This expression is negative for $r\in(r_o,\infty)$. Thus, there are no closed trapped surfaces in the Fisher spacetime. For the Fisher universe \eq{3.2} $\mathcal{T}$ is defined by $\tau=const$ and $\rho=const$ and the trapping scalar is
\be\n{4.4}
\Theta_{ \mathcal{T}}=\left.\frac{(d-2)^2}{8\rho^{S+d-2}}\frac{\left((1+S)\rho_o^{d-3}-2\rho^{d-3}\right)^2}{\left(\rho_o^{d-3}-\rho^{d-3}\right)^{\tfrac{\mathstrut 1-S}{d-3}+1}}\right\vert_{\rho=const.}
\ee
Clearly, it is nonnegative for $\rho\in(0,\rho_o)$. The trapping scalar vanishes for
\be\n{4.5}
\rho=\rho_*=\rho_o\left(\frac{1+S}{2}\right)^{\tfrac{\mathstrut 1}{d-3}}.
\ee
The corresponding spacelike $(d-2)$-dimensional surface is called a {\em marginally trapped surface}. Note that $R_{\hta\hrho\hta}^{\ \ \ \ \hrho}$ is zero on this surface [see, \eq{B1a} and \eq{3.1}]. For the Schwarzschild-Tangherlini black hole the {\em marginally trapped surface} coincides with the surface of its event horizon: $\rho_*=\rho_o=r_o$. In the case of the massless Fisher solution we have $\rho_*^{d-3}=\rho_o^{d-3}/2$. 

\begin{figure}[htb]
\begin{center}
\ba
&\hspace{-0.2cm}\includegraphics[height=3.73cm,width=4.0cm]{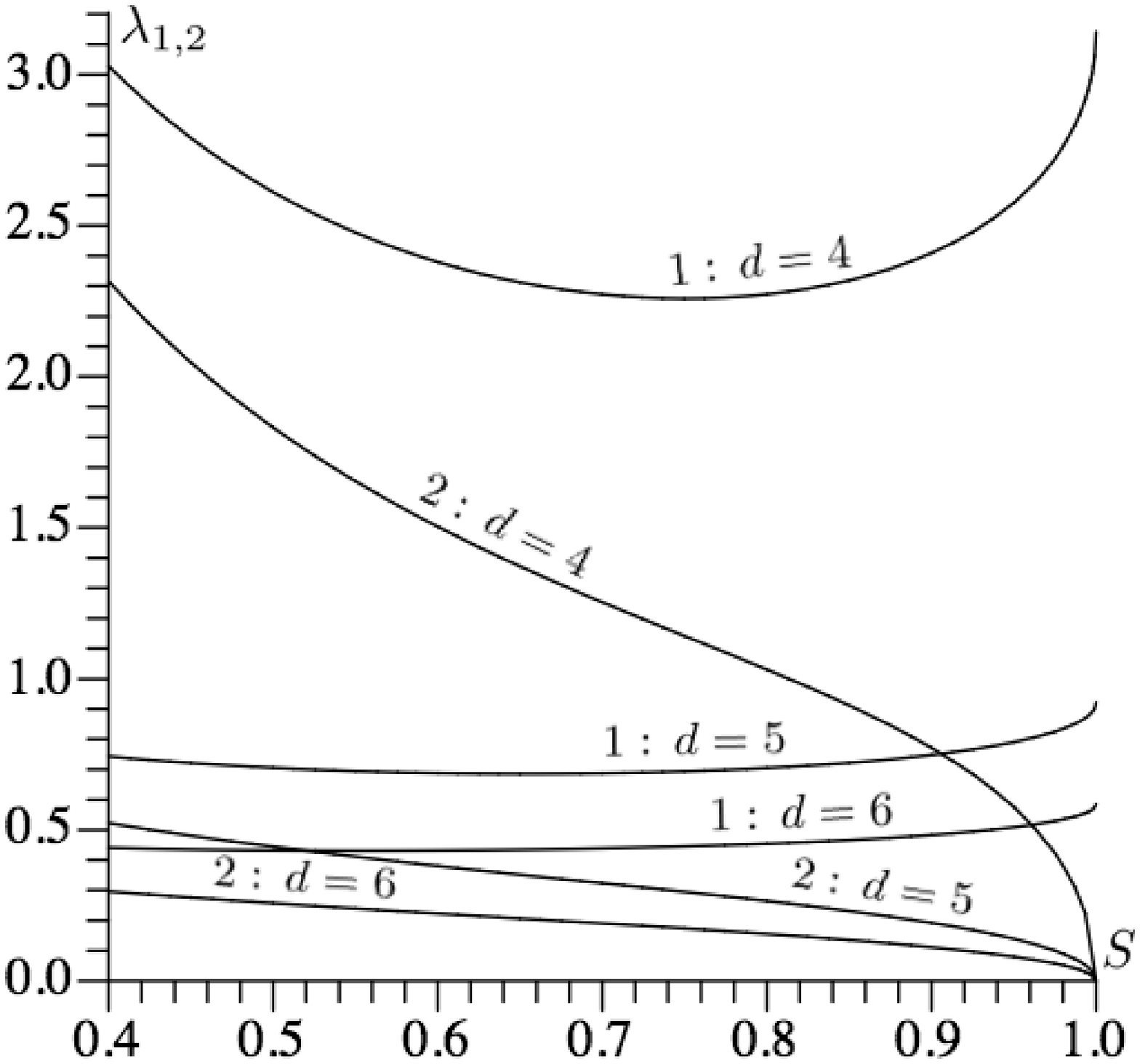}
&\hspace{0.2cm}\includegraphics[height=3.83cm,width=4.0cm]{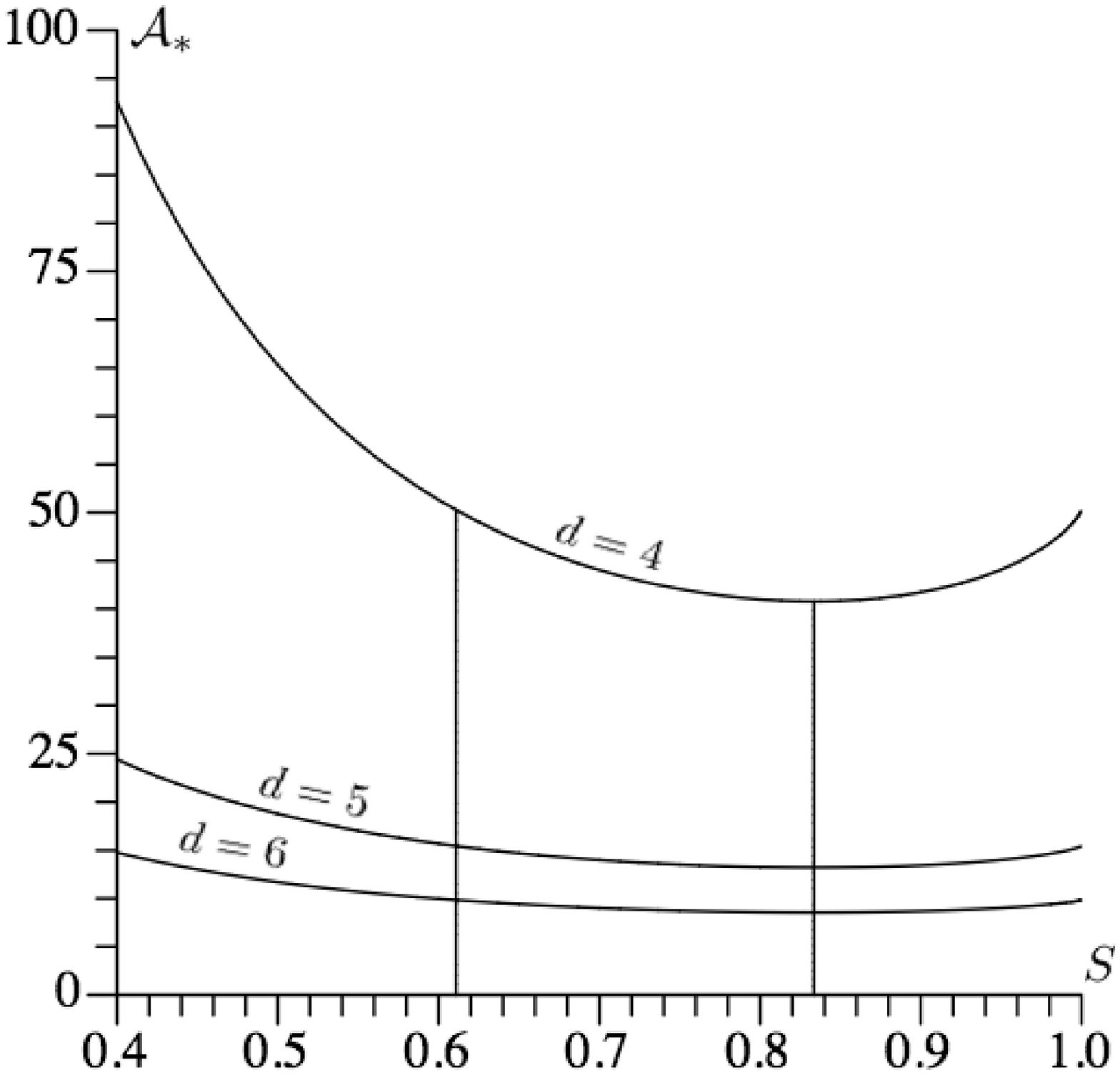}
\nonumber\\
& ({\bf a}) &\hspace{2.1cm}({\bf b})\nonumber\ea
\vspace{-0.5cm}
\caption{({\bf a}): Maximal proper time $\lambda_{1,2}$ as a function of $S$ for the fixed value of the mass $\mu=1$ and $d=4,5,6$. The indices $1$ and $2$ correspond to $\lambda_1$ and $\lambda_2$, respectively. ({\bf b}): Area $\mathcal{A}_{*}$ as a function of $S$ for $\mu=1$ and $d=4,5,6$. In any dimension, the minimal value of $\mathcal{A}_{*}$ corresponds to $S\approx 0.834$ and for $S\approx 0.611$ the value of $\mathcal{A}_{*}$ equals to the horizon surface area of the Schwarzschild-Tangherlini black hole of $\mu=M=1$ and $\Sigma=0$, $(S=1)$.}\label{F1}
\end{center} 
\end{figure}

Let us calculate the maximal proper time $\lambda_1$ corresponding to the interval $\rho\in(0,\rho_*]$ for the radial timelike geodesics. Using \eq{3.19a} and \eq{3.19b} and taking $\mathscr{E}_{-1}=0$ we derive
\ba
\lambda_1&=&\int^{\rho_*}_{+0}d\rho\left[\left(\frac{\rho_o}{\rho}\right)^{d-3}-1\right]^{\tfrac{\mathstrut 1-S}{2(d-3)}-\tfrac{\mathstrut 1}{2}}\non\\
&=&\frac{\rho_o}{d-3}\mathcal{B}_{\tfrac{\mathstrut 1+S}{2}}\left(\frac{1+S}{2(d-3)}+\frac{1}{2},\frac{1-S}{2(d-3)}+\frac{1}{2}\right)\non\,,\\
\ea
where $\mathcal{B}_x(a,b)$ is the incomplete beta function (see, e.g., \cite{AS}, p. 263). The maximal proper time $\lambda_2$ corresponding to the interval $\rho\in[\rho_*,\rho_o)$ is
\ba
\lambda_2&=&\int^{\rho_o-0}_{\rho_*}d\rho\left[\left(\frac{\rho_o}{\rho}\right)^{d-3}-1\right]^{\tfrac{\mathstrut 1-S}{2(d-3)}-\tfrac{\mathstrut 1}{2}}\non\\
&=&\frac{\rho_o}{d-3}\mathcal{B}\left(\frac{1+S}{2(d-3)}+\frac{1}{2},\frac{1-S}{2(d-3)}+\frac{1}{2}\right)-\lambda_1\non\,,\\
\ea
where $\mathcal{B}(a,b)$ is the beta function (see, e.g., \cite{AS}, p. 258). According to the symmetry property of the incomplete beta function,
\be
\mathcal{B}_x(a,b)=\mathcal{B}(a,b)-\mathcal{B}_{1-x}(b,a)\,,
\ee
for the massless Fisher solution $(S=0)$ we have 
\be
\lambda_1=\lambda_2=\frac{\rho_o}{2(d-3)}\mathcal{B}\left(\frac{d-2}{2(d-3)},\frac{d-2}{2(d-3)}\right)\,. 
\ee
For the Schwarzschild-Tangherlini black hole $(S=1)$ we have
\be
\lambda_1=\frac{r_o}{d-3}\mathcal{B}\left(\frac{1}{d-3}+\frac{1}{2},\frac{1}{2}\right)\,,\hhh\lambda_2=0\,.
\ee 
In the 4-dimensional case this expression reduces to the well-known result: $\lambda_1=\pi M$ (see \cite{MTW}, p. 836). Let us see how $\lambda_1$ and $\lambda_2$ depend on the scalar charge $\Sigma$. Figure~\ref{F1}({\bf a}) illustrates the maximal proper time $\lambda_1$ and $\lambda_2$ as a function of $S$ for the fixed value of the mass $\mu=1$. Thus, $S=0$ corresponds to infinite value of the scalar charge $\sigma$ [see expression \eq{3.1a}] and, as a result, $\lambda_1=\lambda_2\to\infty$. Note, that for any $d\geqslant4$ the maximal proper time $\lambda_1$ has a local minimum for a certain value of $S\in(0,1)$.
 
Let us calculate the area of the marginally trapped surface defined by \eq{4.5}. The areal radius corresponding to $\rho_*$ is 
\be\n{4.10}
\mathscr{R}_{*}\equiv\mathscr{R}(\rho_*)=2^{-\tfrac{\mathstrut 1}{d-3}}\rho_o(1-S)^{\tfrac{\mathstrut1-S}{2(d-3)}}(1+S)^{\tfrac{\mathstrut1+S}{2(d-3)}}.
\ee
For $S=1$ we have $\mathscr{R}_{*}=\rho_o=r_o$, which corresponds to the Schwarzschild-Tangherlini black hole, and for $S=0$ we have $\mathscr{R}_{*}=2^{-\tfrac{\mathstrut 1}{d-3}}\rho_o$, which corresponds to the massless Fisher solution. Thus, the area of the $(d-2)$-dimensional marginally trapped surface is
\be\n{4.10a}
\mathcal{A}_{*}=\frac{2\mathscr{R}_{*}^{d-2}\pi^{\frac{d-1}{2}}}{\Gamma(\frac{d-1}{2})}.
\ee
Figure~\ref{F1}({\bf b}) illustrates how this area depends on the value of $S$ for the fixed value of the mass $\mu=1$. 

\subsection{Misner-Sharp energy}

In a spherically symmetric spacetime the Misner-Sharp energy $M(R)$, which is a spacetime invariant, defines the ``local gravitational energy'' inside a sphere of the areal radius $R$ (see, e.g.,\cite{MTW,MS}). It has many interesting properties (see, e.g., \cite{Hay}). In particular, at spatial infinity in an asymptotically flat spacetime it reduces to the Arnowitt-Deser-Misner energy. For a central singularity, a negative value of the Misner-Sharp energy implies that the singularity is untrapped and timelike. If the dominant energy condition holds on an untrapped sphere, the Misner-Sharp energy is monotonically increasing in outgoing spatial or null directions. As we shall see below, this is exactly the case for the central singularity at $r=r_o$ of the Fisher spacetime. Here we use the following expression for the Misner-Sharp energy generalized to a $d$-dimensional spacetime: 
\be\n{4.5a}
M(r)=\frac{(d-2)\pi^{\frac{d-3}{2}}}{8\Gamma(\frac{d-1}{2})}R(r)^{d-3}\left[1-g^{rr}(r)\left(\frac{dR(r)}{dr}\right)^2\right]\,.
\ee
For the Fisher spacetime \eq{2.4}, $r\in(r_o,\infty)$ we have
\be\n{areal}
R(r)=r\left[1-\left(\frac{r_o}{r}\right)^{d-3}\right]^{\tfrac{\mathstrut 1-S}{2(d-3)}}\,
\ee
and the Misner-Sharp energy is
\ba\n{4.5b}
M(r)&=&\frac{(d-2)\pi^{\frac{d-3}{2}}}{32\Gamma(\frac{d-1}{2})}\frac{r_o^{d-3}\left(4Sr^{d-3}-(1+S)^2r_o^{d-3}\right)}{r^{(1-S)\tfrac{d-3}{2}}\left(r^{d-3}-r_o^{d-3}\right)^{\tfrac{\mathstrut 1+S}{2}}}\,.\non\\
\ea
In the limit $r\to\infty$ we have $M(r)\to M$. The Misner-Sharp energy \eq{4.5b} vanishes for
\be\n{4.5c}
r=r_e=r_o(4S)^{-\tfrac{\mathstrut 1}{d-3}}(1+S)^{\tfrac{\mathstrut 2}{d-3}}\,,\hhh r_e>r_o\,,
\ee
where $S\in[0,1)$, and it is negative for $r\in(r_o,r_e)$. Note that $R_{\hal\hbe\hal}^{\ \ \ \ \hbe}$ is zero for $r=r_e$ [see, \eq{B1d}]. For $S=1$, which corresponds to the Schwarzschild-Tangherlini black hole, we have 
\be
M(r)=\frac{(d-2)\pi^{\frac{d-3}{2}}}{8\Gamma(\frac{d-1}{2})}r_o^{d-3}=M\geqslant0\,.
\ee
For a negative mass Schwarzschild-Tangherlini spacetime which has naked singularity, $M(r)=M<0$ everywhere. 

\begin{figure}[htb]
\begin{center}
\includegraphics[height=5.15cm,width=6.0cm]{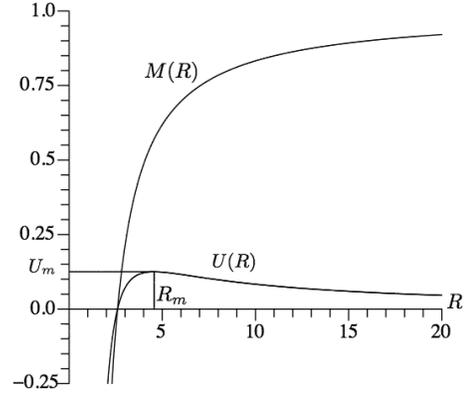}
\caption{Misner-Sharp energy $M(R)$ and the ``local (Newtonian) gravitational potential energy" $U(R)$ for $M=1$, $S=1/2$, and $d=4$. The maximum $U_m=U(R_m)$ corresponds to $R_m=2\cdot3^{3/4}$ and is equal to 1/8, \cite{RN}.}\label{F2}
\end{center} 
\end{figure}

The Misner-Sharp energy \eq{4.5a} can be expressed in terms of the trapping scalar $\Theta_{\mathcal{T}}$ [see, \eq{4.3}] as follows:
\be\n{4.5d}
M(r)=\frac{(d-2)\pi^{\frac{d-3}{2}}}{8\Gamma(\frac{d-1}{2})}R(r)^{d-3}\left[1+\frac{2R(r)^2}{(d-2)^2}\Theta_{\mathcal{T}}\right]\,.
\ee
Thus, it defines a condition when a sphere of the areal radius $R$ is trapped. Another way to define this condition is to introduce the ``local (Newtonian) gravitational potential energy" associated with $M(r)$ as follows:
\be
U(R)=\frac{4\Gamma(\frac{d-1}{2})M(r)}{(d-2)\pi^{\frac{d-3}{2}}R(r)^{d-3}}\,.
\ee
Then, the trapping condition is the following: if $U(R)>1/2$ the surface $R=const$ is trapped, if $U(R)=1/2$ the surface is marginally trapped, and if $U(R)<1/2$ the surface is untrapped. Figure~\ref{F2} illustrates $M(R)$ and $U(R)$ for $d=4$. For any $d\geqslant4$, $M(R)$ is monotonically increasing and $U(R)$ has the maximum $U_m=U(R_m)=S^2/2\leqslant1/2$, where
\be\n{Rm}
R_m\equiv R(r_m)=r_o(2S)^{-\tfrac{\mathstrut 1}{d-3}}(1-S)^{\tfrac{\mathstrut 1-S}{2(d-3)}}(1+S)^{\tfrac{\mathstrut 1+S}{2(d-3)}}\,,
\ee 
and
\be\n{4.5e}
r_m=r_o\left(\frac{1+S}{2S}\right)^{\tfrac{\mathstrut 1}{d-3}}\,,\hhh r_m\geqslant r_e\,.
\ee
For the Schwarzschild-Tangherlini black hole we have $R_m=r_o$. Note that $R_{\hr\hal\hr}^{\ \ \ \ \hal}$ is zero for $r=r_m$ [see, \eq{B1c}].

\begin{figure}[htb]
\begin{center}
\ba
&\hspace{-0.2cm}\includegraphics[height=3.86cm,width=4.0cm]{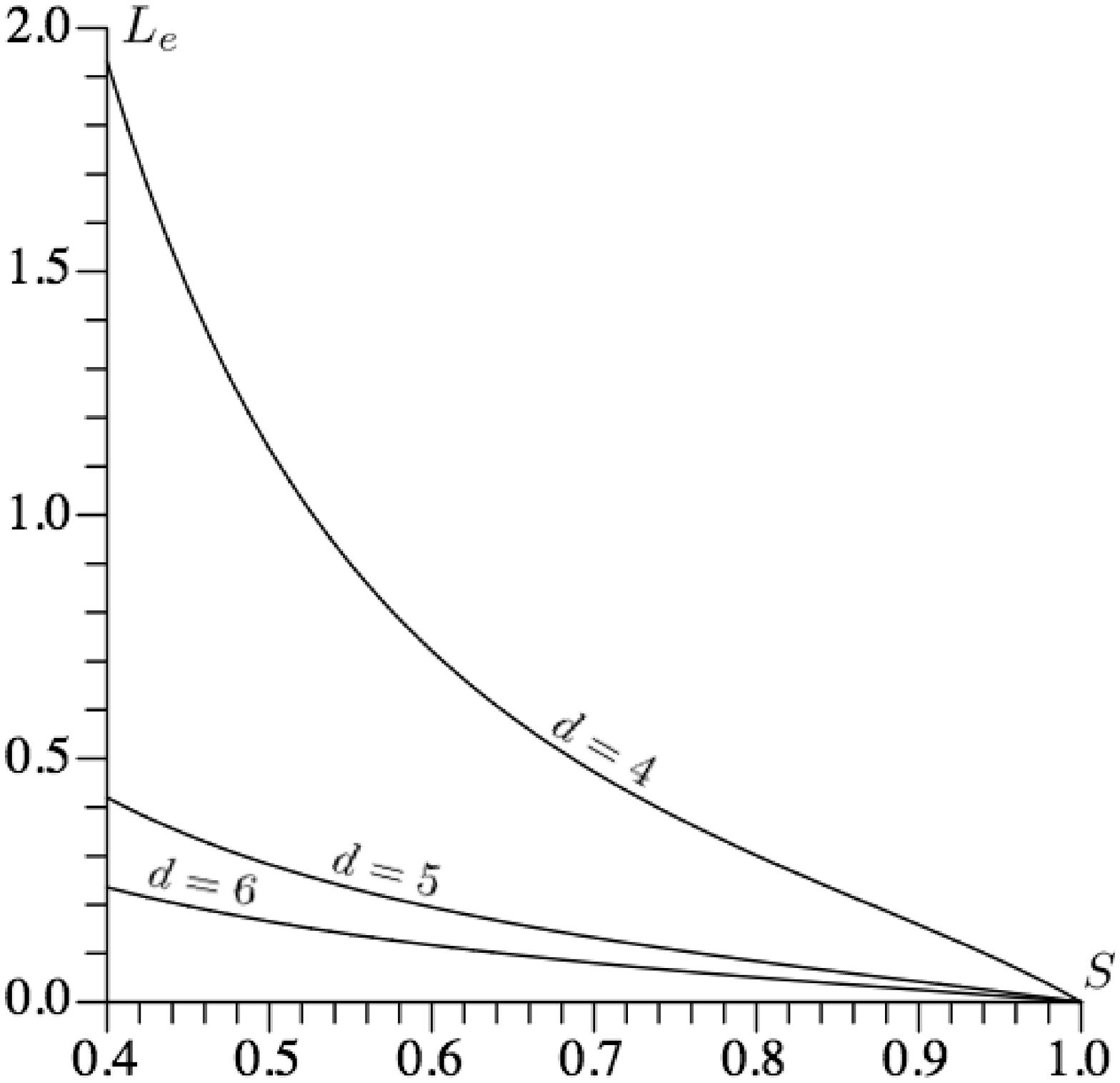}
&\hspace{0.2cm}\includegraphics[height=3.83cm,width=4.0cm]{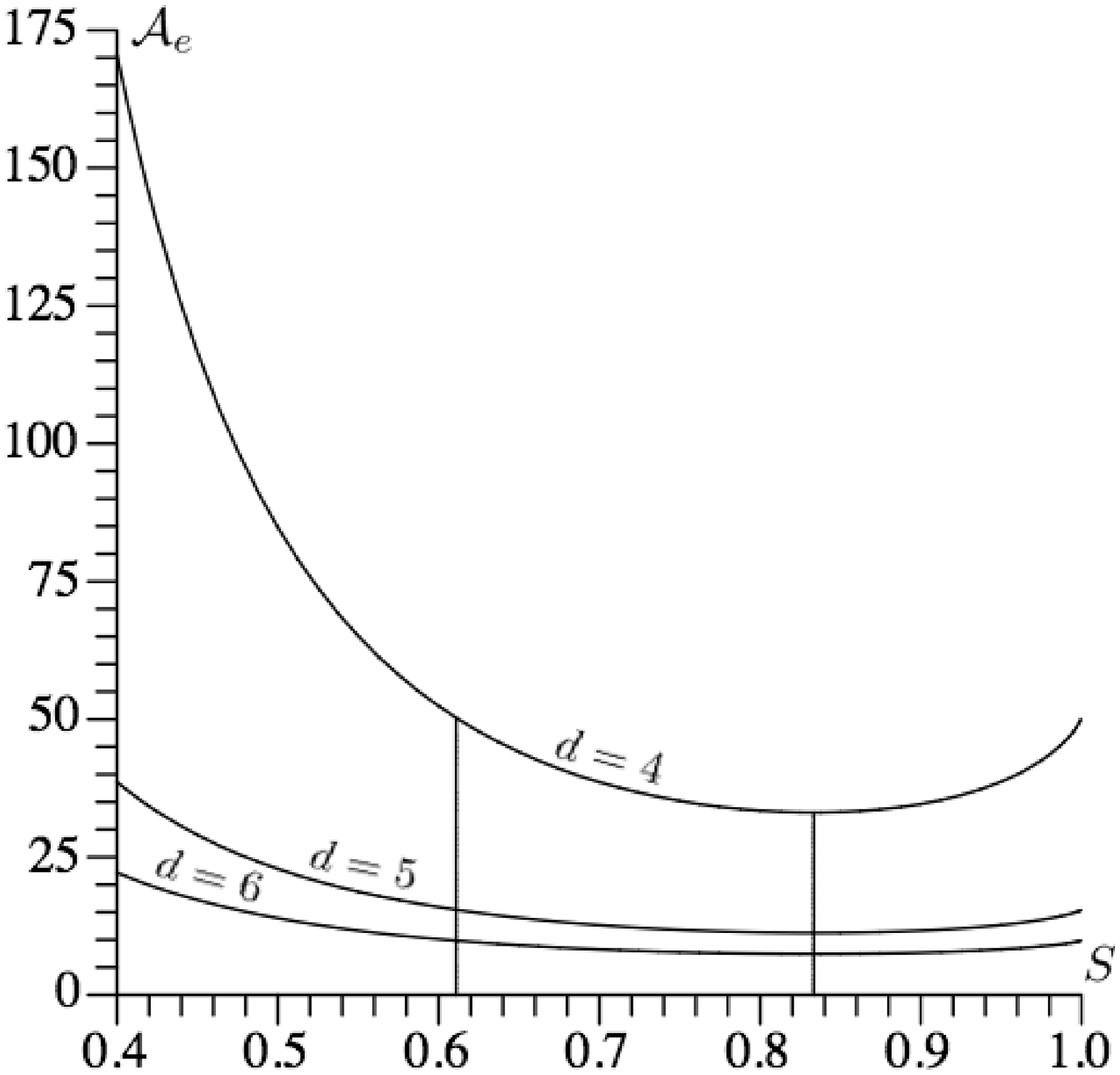}
\nonumber\\
& ({\bf a}) &\hspace{2.1cm}({\bf b})\nonumber
\ea
\caption{({\bf a}): Proper distance $L_{e}$ as a function of $S$ for the fixed value of the mass $M=1$ and $d=4,5,6$. ({\bf b}): Area $\mathcal{A}_{e}$ as a function of $S$ for $M=1$. In any dimension, the minimal value of $\mathcal{A}_{e}$ corresponds to $S\approx 0.834$ and for $S\approx 0.611$ the value of $\mathcal{A}_{e}$ equals to the horizon surface area of the Schwarzschild-Tangherlini black hole of $M=1$ and $\Sigma=0$, $(S=1)$.}\label{F3}
\end{center} 
\end{figure}

Let us calculate geometric invariants of the region where $M(r)\leqslant 0$. The proper distance corresponding to nonpositive $M(r)$ is
\be
L_e=\int^{r_e}_{r_o+0} dr\left[1-\left(\frac{r_o}{r}\right)^{d-3}\right]^{\tfrac{\mathstrut 1-S}{2(d-3)}-\tfrac{\mathstrut 1}{2}}\,.
\ee
Figure~\ref{F3}({\bf a}) illustrates the proper distance $L_e$ as a function of $S$ for the fixed value of the mass $M=1$. According to the figure, the proper distance $L_e$ is a monotonically decreasing function of $S$. This function diverges for $S\to0$ corresponding to infinite value of the scalar charge $\Sigma$ [see expression \eq{2.6a}]. 
 
Let us calculate the area of the sphere corresponding to zero Misner-Sharp energy. The areal radius corresponding to $r_e$ [see, \eq{4.5c}] is
\be\n{Re}
R_e\equiv R(r_e)=r_o(4S)^{-\tfrac{\mathstrut 1}{d-3}}(1-S)^{\tfrac{\mathstrut 1-S}{d-3}}(1+S)^{\tfrac{\mathstrut 1+S}{d-3}}\,.
\ee
For the Schwarzschild-Tangherlini black hole $(S=1)$ we have $R_e=r_o$ and for the massless Fisher solution $(S=0)$ we have $R_e\to +\infty$. The area of the $(d-2)$-dimensional sphere corresponding to zero Misner-Sharp energy is
\be\n{Ae}
\mathcal{A}_{e}=\frac{2R_e^{d-2}\pi^{\frac{d-1}{2}}}{\Gamma(\frac{d-1}{2})}\,.
\ee
Figure~\ref{F3}({\bf b}) illustrates how this area depends on the value of $S$ for the fixed value of the mass $M=1$. It is remarkable that in any dimension $d\geqslant4$ both the areas $\mathcal{A}_{e}$ and $\mathcal{A}_{*}$ [see Fig.~\ref{F1}({\bf b})] have minimal values at the same value of $S\approx 0.834$, and for $S\approx 0.611$ they are equal to the horizon surface area of the Schwarzschild-Tangherlini black hole of $M=1$ and $\Sigma=0$, $(S=1)$.

\subsection{Causal structure}

To study the causal structure of the Fisher spacetime and the Fisher universe we consider first radial null geodesics. We start from the Fisher spacetime \eq{2.4}, $r\in(r_o,\infty)$ and consider radial null geodesics in the $(t,R)$ plane, where $R=R(r)$ is the areal radius [see, \eq{areal}], which is a geometric invariant. Using \eq{3.12a} and \eq{3.12b} we present the solution for the radial null geodesics in the following form:
\ba
t(r)&=&\pm\int dr\left[1-\left(\frac{r_o}{r}\right)^{d-3}\right]^{\tfrac{\mathstrut 1-S}{2(d-3)}-\tfrac{\mathstrut 1+S}{2}}\,,\n{4.6a}\\
R(r)&=&r\left[1-\left(\frac{r_o}{r}\right)^{d-3}\right]^{\tfrac{\mathstrut 1-S}{2(d-3)}}\,,\n{4.6b}
\ea
where ``$+$" stands for outgoing and ``$-$" stands for ingoing radial null geodesics. The coordinate $t$ is timelike and the areal radius $R(r)$ is spacelike. Local null cones are defined by
\be\n{4.7}
\frac{dt}{dR}=\pm2r^{(1+S)\tfrac{d-3}{2}}\frac{\left(r^{d-3}-r_o^{d-3}\right)^{\tfrac{\mathstrut1-S}{2}}}{2r^{d-3}-(1+S)r_o^{d-3}}.
\ee

\begin{figure}[htb]
\begin{center} 
\includegraphics[height=5.99cm,width=6cm]{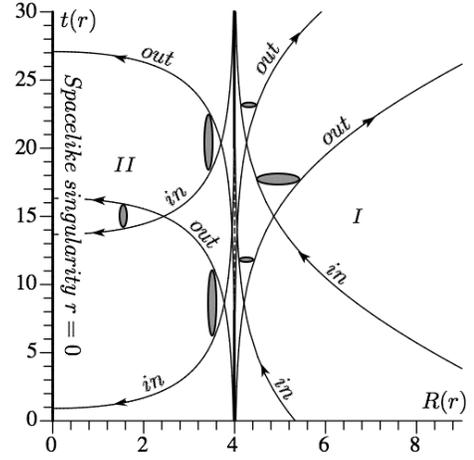} 
\caption{Radial null geodesics in the Schwarzschild-Tangherlini spacetime of $M'=2$, $\Sigma'=0$, [see, \eq{st}] and $d=4$. The behavior of the geodesics is generic for other values of $d>4$. The black hole event horizon is located at $R=r_o=4$. It separates the exterior $I$ and interior $II$ regions. The spacelike singularity is located at $R=r=0$. The direction of local time is illustrated by the future null cones.}\label{F4} 
\end{center}
\end{figure}
\begin{figure}[htb]
\begin{center} 
\includegraphics[height=6.0cm,width=6.0cm]{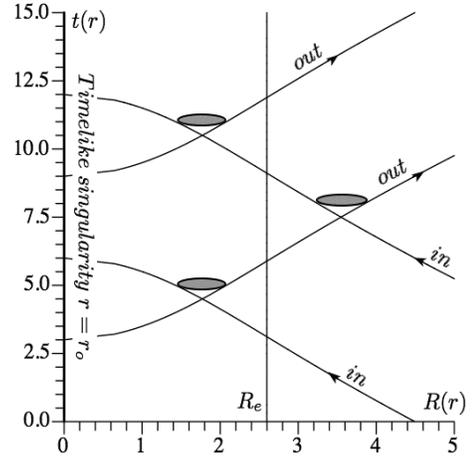}
\caption{Radial null geodesics in the Fisher spacetime of $M=1$, $S=1/2$, and $d=4$. The behavior of the geodesics is generic for other values of $d>4$ and $S\in[0,1)$. The areal radius corresponding to zero value of the Misner-Sharp energy is given by $R_e=3\sqrt{3}/2$. The timelike singularity is located at $R(r_o)=0$. The direction of local time is illustrated by the future null cones.}\label{F5a} 
\end{center}
\end{figure}
\begin{figure}[htb]
\begin{center} 
\includegraphics[height=3.91cm,width=8.0cm]{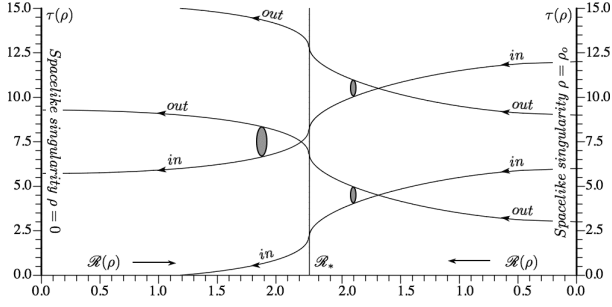}
\caption{Radial null geodesics in the Fisher universe of  $\mu=1$, $S=1/2$, and $d=4$. The behavior of the geodesics is generic for other values of $d>4$ and $S\in[0,1)$. The marginally trapped surface is located at $\mathscr{R}_{*}=3^{3/4}$. The spacelike singularities corresponding to $\rho=\rho_o$ and $\rho=0$ are located at $\mathscr{R}=0$. The direction of local time is illustrated by the future null cones.}\label{F5b} 
\end{center}
\end{figure}
\begin{figure}[htb]
\begin{center}
\includegraphics[height=3.34cm,width=7cm]{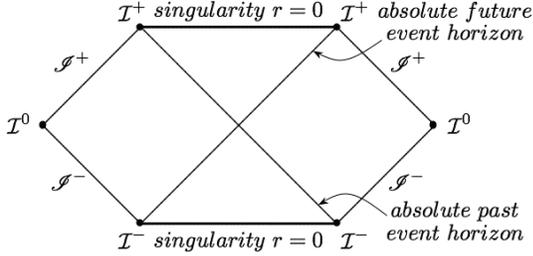}
\caption{Penrose diagram for the maximally extended Schwarzschild-Tangherlini spacetime. Each interior point in the diagram represents a $(d-2)$-dimensional sphere.}\label{D2}
\end{center} 
\end{figure}
\begin{figure}[htb]
\begin{center}
\includegraphics[height=5.6cm,width=5cm]{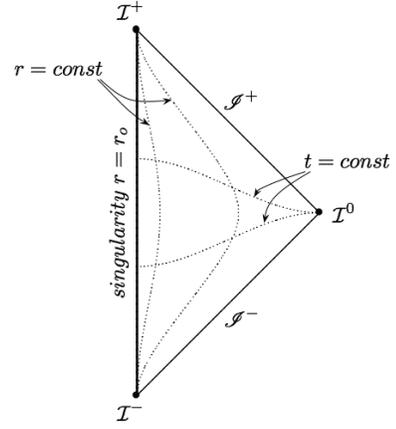}
\caption{Penrose diagram for the Fisher spacetime. Each interior point in the diagram represents a $(d-2)$-dimensional sphere.}\label{D3}
\end{center} 
\end{figure}
\begin{figure}[htb]
\begin{center}
\includegraphics[height=2.9cm,width=8cm]{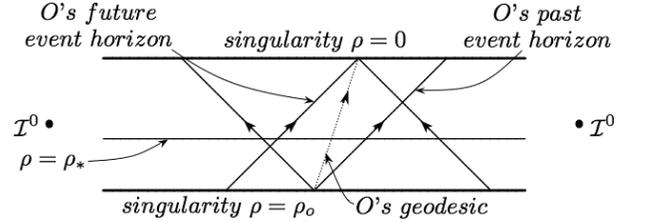}
\caption{Penrose diagram for the Fisher universe. Each interior point in the diagram represents a $(d-2)$-dimensional sphere. The marginally trapped surface of the Fisher universe is schematically illustrated by the infinite line $\rho=\rho_*$.}\label{D4}
\end{center} 
\end{figure}

The radial null geodesics in the Fisher universe \eq{3.2} can be derived by applying the transformations \eq{3.1} to expressions \eq{4.6a} and \eq{4.6b}, or directly by using \eq{3.19a} and \eq{3.19b},
\ba
\tau(\rho)&=&\mp\int d\rho\left[\left(\frac{\rho_o}{\rho}\right)^{d-3}-1\right]^{\tfrac{\mathstrut 1-S}{2(d-3)}-\tfrac{\mathstrut 1+S}{2}}\,,\n{4.8a}\\
\mathscr{R}(\rho)&=&\rho\left[\left(\frac{\rho_o}{\rho}\right)^{d-3}-1\right]^{\tfrac{\mathstrut 1-S}{2(d-3)}}\,,\n{4.8b}
\ea
where ``$-$" stands for outgoing and ``$+$" stands for ingoing radial null geodesics. The coordinate $\tau$ is spacelike and the areal radius $\mathscr{R}(\rho)$ is timelike. The local null cones are defined by 
\be\n{4.9}
\frac{d\mathscr{R}}{d\tau}=\mp\frac{1}{2}\rho^{-(1+S)\tfrac{d-3}{2}}\frac{(1+S)\rho_o^{d-3}-2\rho^{d-3}}{\left(\rho_o^{d-3}-\rho^{d-3}\right)^{\tfrac{\mathstrut1-S}{2}}}.
\ee
This expression vanishes at $\rho=\rho_*$ which corresponds to the marginally trapped surface \eq{4.5}.

The radial null geodesics corresponding to $S=1$ are illustrated in Fig.~\ref{F4}. To construct a similar picture for the radial null geodesics corresponding to $S\in[0,1)$ we define the direction of time in the Fisher universe in accordance with the Schwarzschild-Tangherlini black hole interior (see region $II$ in Fig.~\ref{F4}). Namely, for $S=1$ the timelike coordinate $\rho=\rho_o=r_o$ is past and $\rho=r=0$ is future. We shall keep this convention for other values of $S\in[0,1)$. The radial null geodesics in the Fisher spacetime and the Fisher universe are illustrated in Figs.~\ref{F5a} and \ref{F5b}, respectively.

The Fisher universe is an anisotropic universe whose topology is $\mathbb{R}_{\tau}^1\times \mathbb{R}_{\rho}^1\times \mathbb{S}^{d-2}$. At the moment of its ``Big Bang" $(\rho=\rho_o)$ the Fisher universe is a point of zero proper $(d-1)$-dimensional volume. It begins to expand in all spatial directions and at the moment $\rho=\rho_*$ [see, \eq{4.5}] its boundary area along the angular directions reaches the maximal value $\mathcal{A}_*$ [see, \eq{4.10a}], and the universe begins to contract in the angular directions and continues to expand in the spatial $\tau$ direction. At the moment of its ``Big Crunch" $(\rho=0)$ its boundary area along the angular directions vanishes and its expansion along the $\tau$ direction diverges.  

The causal structure of the Fisher solution can be summarized in the corresponding Penrose diagrams (see Figs.~\ref{D3} and \ref{D4}). For comparison, we present the Penrose diagram of the Schwarzschild-Tangherlini spacetime (see Fig.~\ref{D2}). The topology of the spacelike singularity located at $r=0$ is $\mathbb{R}_{t}^1\times \mathbb{S}^{d-2}$ \cite{TOP}. Figure~\ref{D3} represents the region conformal to the Fisher spacetime \eq{2.4}, $r\in(r_o,\infty)$. It is asymptotically flat and has timelike curvature singularity at $r=r_o$. The topology of the timelike singularity located at $r=r_o$ is $\mathbb{R}_{t}^1$ for $S\in[0,1/(d-2))$, and $\mathbb{R}_{t}^1\times \mathbb{S}^{d-2}$ for $S\in[1/(d-2),1)$. Figure~\ref{D4} represents the region conformal to the Fisher universe \eq{3.2}. The coordinate $\rho$ and the corresponding ``tortoise coordinate," which is given by the right-hand side of \eq{4.8a}, take finite values, whereas $\tau\in(-\infty,\infty)$. There is no conformal transformation which makes the infinite interval $\tau\in(-\infty,\infty)$ finite and does not shrink the finite interval of the tortoise coordinate to a point, thus inducing a coordinate singularity \cite{AdS}. Here we present spacelike infinities $\tau\to\pm\infty$ by two disjoint points $\mathcal{I}^0$. The spacetime singularities of the Fisher universe located at $\rho=\rho_o$ and $\rho=0$ are both spacelike. The topology of the spacelike singularities located at $\rho=\rho_o$ and at $\rho=0$ is $\mathbb{R}_{\tau}^1\times \mathbb{S}^{d-2}$. According to the time direction convention the singularity at $\rho=\rho_o$ is in the past and the singularity $\rho=0$ is in future. Thus, any causal curve in the Fisher universe originates at $\rho=\rho_o$ and terminates at $\rho=0$. As a result, for geodesic families of observers both particle and event horizons exist. The geodesic of one such observer $O$ and the corresponding past and future event horizons are shown in the diagram.  

\section{Isometric embedding}

One of the ways to study geometry of a $d$-dimensional (pseudo-)Riemannian space which has an analytic metric of signature $p-q\leqslant d$ is to construct its isometric embedding into a $D$-dimensional (pseudo-)Euclidean space with the signature $r-s\leqslant D$. A local analytic isometric embedding is always possible if the dimension of the (pseudo-)Euclidean space of the signature $r-s$ is $D=d(d+1)/2$ and $r\geqslant p,s\geqslant q\,$, \cite{Fri}. For a global isometric embedding the dimension $D$ generally should be greater \cite{Nash}. For example, a 4-dimensional Schwarzschild solution whose metric has the signature $3-1=2$ can be isometrically embedded into a 6-dimensional pseudo-Euclidean space of the signature $5-1=4$ \cite{Fro}. Examples of isometric local and sometimes global embeddings of some 4-dimensional Lorentzian spacetimes  into pseudo-Euclidean spaces of higher dimensions are given in \cite{Ros}. When dealing with spacetimes of the general theory of relativity one has usually $d\geqslant 4$ and higher values of $D$. Thus, having an embedding it is impossible to construct the corresponding visual picture illustrating the spacetime geometry. However, if a spacetime has symmetries defined by its Killing vectors, one can study its geometry by considering embeddings of the spacetime (hyper)surfaces orthogonal to the orbits of its Killing vectors. In the case if such a 2-dimensional surface exists, one can construct a 3-dimensional picture illustrating its isometric local embedding. 

Here we shall consider local isometric embeddings of 2-dimensional subspaces of the Fisher spacetime and the Fisher universe. Both the spacetimes have a set of Killing vectors which allows us to study their geometry by considering embedding of the corresponding 2-dimensional subspaces. The geometry of the Fisher spacetime \eq{2.4}, $r\in(r_o,\infty)$ and the Fisher universe \eq{3.2} is the same for any value of the coordinate $t$ and $\tau$, respectively. In addition, the spacetimes spherical symmetry implies that any 2-dimensional surface defined by $t=const$ $(\tau=const)$ and $\theta^{\alpha}=const$, $\alpha=3,...,d-1$, where $\theta^{\alpha}\in[0,\pi]$ and $\phi\in[0,2\pi)$ are $d-2$ (hyper)spherical coordinates, has the same geometry. Thus, to visualize the geometry of the spacetimes we present local isometric embeddings of their 2-dimensional subspaces defined by $t=const$ $(\tau=const)$ and $\theta^{\alpha}=\pi/2$, $\alpha=3,...,d-1$. 
  
Let us begin with the Fisher spacetime \eq{2.4} whose 2-dimensional subspace metric is given by
\be\n{5.1}
ds^2=F^{\tfrac{\mathstrut 1-S}{d-3}-1}dr^2+r^2F^{\tfrac{\mathstrut 1-S}{d-3}}d\phi^2,
\ee
where $r\in(r_o,\infty)$ and $F$ is given by \eq{2.4a}. Let us embed this surface into a 3-dimensional Euclidean space endowed with the following metric:
\be\n{5.2}
dl^2=dZ^2+dR^2+R^2d\phi^2,
\ee
where $(Z,R,\phi)$ are the cylindrical coordinates. To construct the embedding we consider the following parametrization of the surface:
\be\n{5.2a}
Z=Z(r),\hhh R=R(r).
\ee
Thus, the surface metric in the cylindrical coordinates takes the following form:
\be\n{5.2b}
dl^2=(Z_{,r}^2+R_{,r}^2)dr^2+R(r)^2d\phi^2.
\ee
Matching the metrics \eq{5.1} and \eq{5.2b} we derive the following embedding map:
\ba
R(r)&=&r\left[1-\left(\frac{r_o}{r}\right)^{d-3}\right]^{\tfrac{\mathstrut 1-S}{2(d-3)}}\,,\n{5.3a}\\
Z(r)&=&r_o^{\tfrac{\mathstrut d-3}{2}}\int dr\hhhhh\frac{[4Sr^{d-3}-(1+S)^2r_o^{d-3}]^{\tfrac{1}{2}}}{2r^{\tfrac{\mathstrut 1-S}{2}}(r^{d-3}-r_o^{d-3})^{1-\tfrac{\mathstrut 1-S}{2(d-3)}}}\,.\non\\\n{5.3b}
\ea
We see that for $r\in(r_o,r_e)$, where $r_e$ is given by \eq{4.5c}, the coordinate $Z(r)$ is imaginary. Thus, the corresponding region of the surface cannot be isometrically embedded in this way into the 3-dimensional Euclidean space. Note that the Misner-Sharp energy \eq{4.5b} and $R_{\hal\hbe\hal}^{\ \ \ \ \hbe}$ \eq{B1d} are negative in this region.

\begin{figure}[htb]
\begin{center}
\ba
&\hspace{-0.25cm}\includegraphics[height=2.08cm,width=4cm]{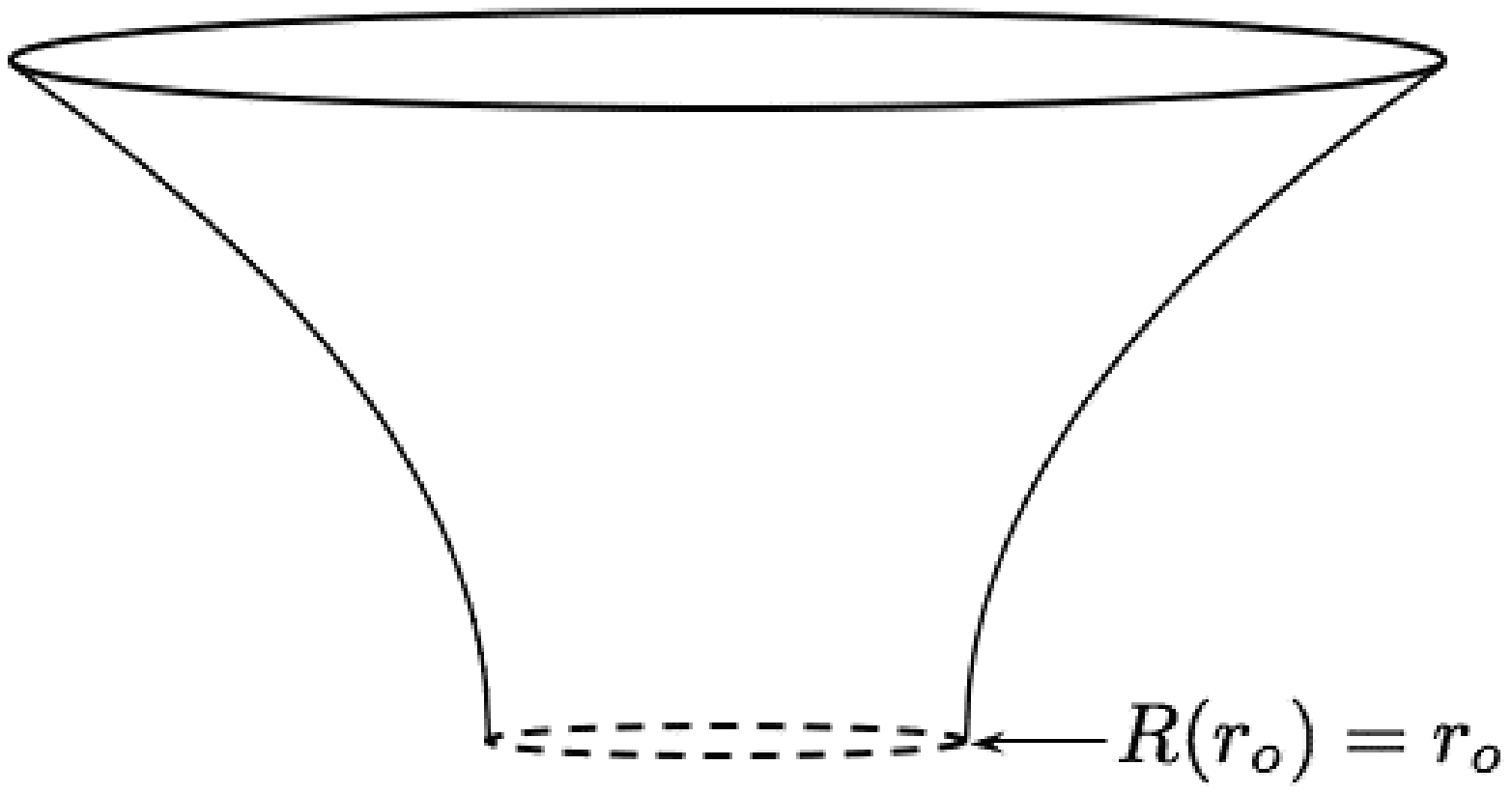}
&\hspace{0.25cm}\includegraphics[height=1.25cm,width=4cm]{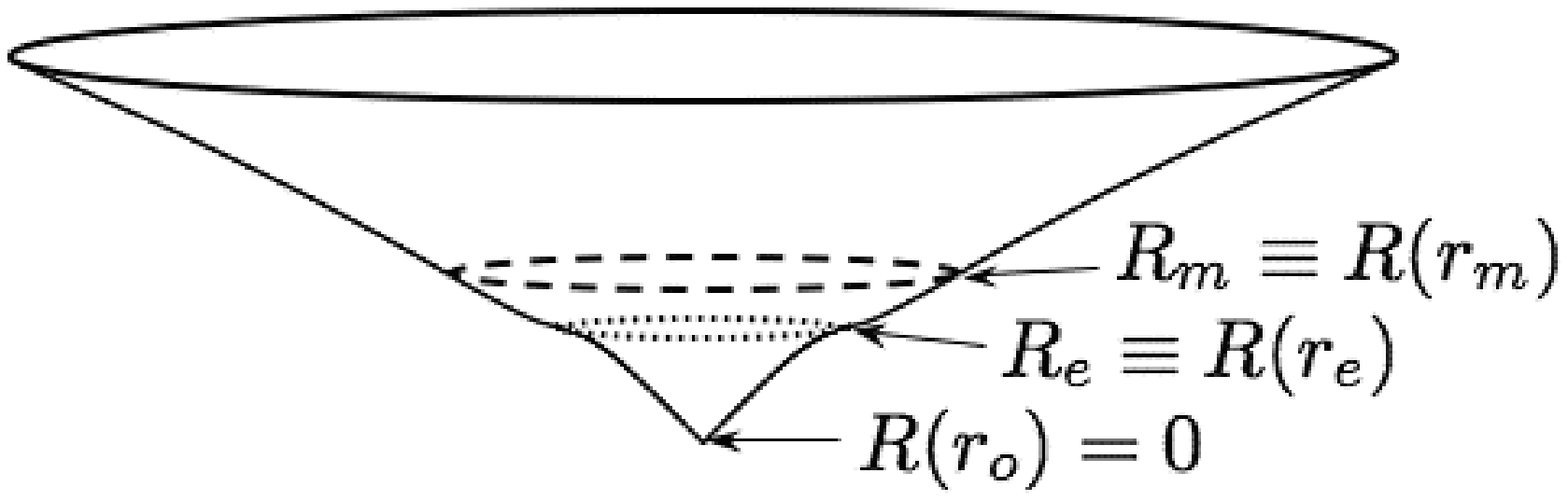}
\nonumber\\
&\hspace{-0.4cm} ({\bf a}) &\hspace{1.8cm}({\bf b})\nonumber
\ea
\vspace{-0.5cm}
\caption{Embedding diagrams for $d=4$. ({\bf a}): Exterior region of the Schwarzschild-Tangherlini black hole of $M'=2$ and $\Sigma'=0$, [see, \eq{st}]. The dashed circle of the radius $R(r_o)=r_o$ represents its event horizon. ({\bf b}): Fisher spacetime corresponding to $M=1$ and $S=1/2$. The point $R(r_o)=0$ represents the naked timelike singularity. The dotted circle of the radius $R_e$ represents the region where the Misner-Sharp energy is zero. The dashed circle of the radius $R_m$ represents the region where the ``local (Newtonian) gravitational potential energy" $U(R)$ is minimal (see Fig.~\ref{F2}). The diagrams are qualitatively generic for other values of $d>4$.}\label{F6}
\end{center} 
\end{figure}

Although the region $r\in(r_o,r_e)$ cannot be isometrically embedded in this way into the 3-dimensional Euclidean space, we can embed it isometrically into 3-dimensional pseudo-Euclidean space endowed with the following metric:
\be\n{5.6}
dl^2=-d\mathcal{Z}^2+dR^2+R^2d\phi^2,
\ee 
where $\mathcal{Z}$ is a timelike coordinate. Repeating the steps above we derive the corresponding embedding map
\ba
R(r)&=&r\left[1-\left(\frac{r_o}{r}\right)^{d-3}\right]^{\tfrac{\mathstrut 1-S}{2(d-3)}}\,,\n{5.7a}\\
\mathcal{Z}(r)&=&r_o^{\tfrac{\mathstrut d-3}{2}}\int dr\hhhhh\frac{[(1+S)^2r_o^{d-3}-4Sr^{d-3}]^{\tfrac{1}{2}}}{2r^{\tfrac{\mathstrut 1-S}{2}}(r^{d-3}-r_o^{d-3})^{1-\tfrac{\mathstrut 1-S}{2(d-3)}}}.\non\\\n{5.7b}
\ea
Embeddings of the surfaces corresponding to $S=1$ and $S=1/2$ are presented in Figs.~\ref{F6}({\bf a}) and \ref{F6}({\bf b}), respectively. In the case of the Fisher spacetime, the region between $R_e$ [see, \eq{Re}] and asymptotic infinity $(R\to\infty)$ corresponds to positive Misner-Sharp energy. The region between $R(r_o)$ and $R_e$ corresponds to negative Misner-Sharp energy. At the convolution point $R_m$ [see, \eq{Rm}] we have $Z_{,R}=S/\sqrt{1-S^2}$. For $S=0$ we have $R_e\to \infty$ and the Misner-Sharp energy is negative everywhere.

\begin{figure}[htb]
\begin{center}
\ba
&\hspace{-0.25cm}\includegraphics[height=3.05cm,width=4cm]{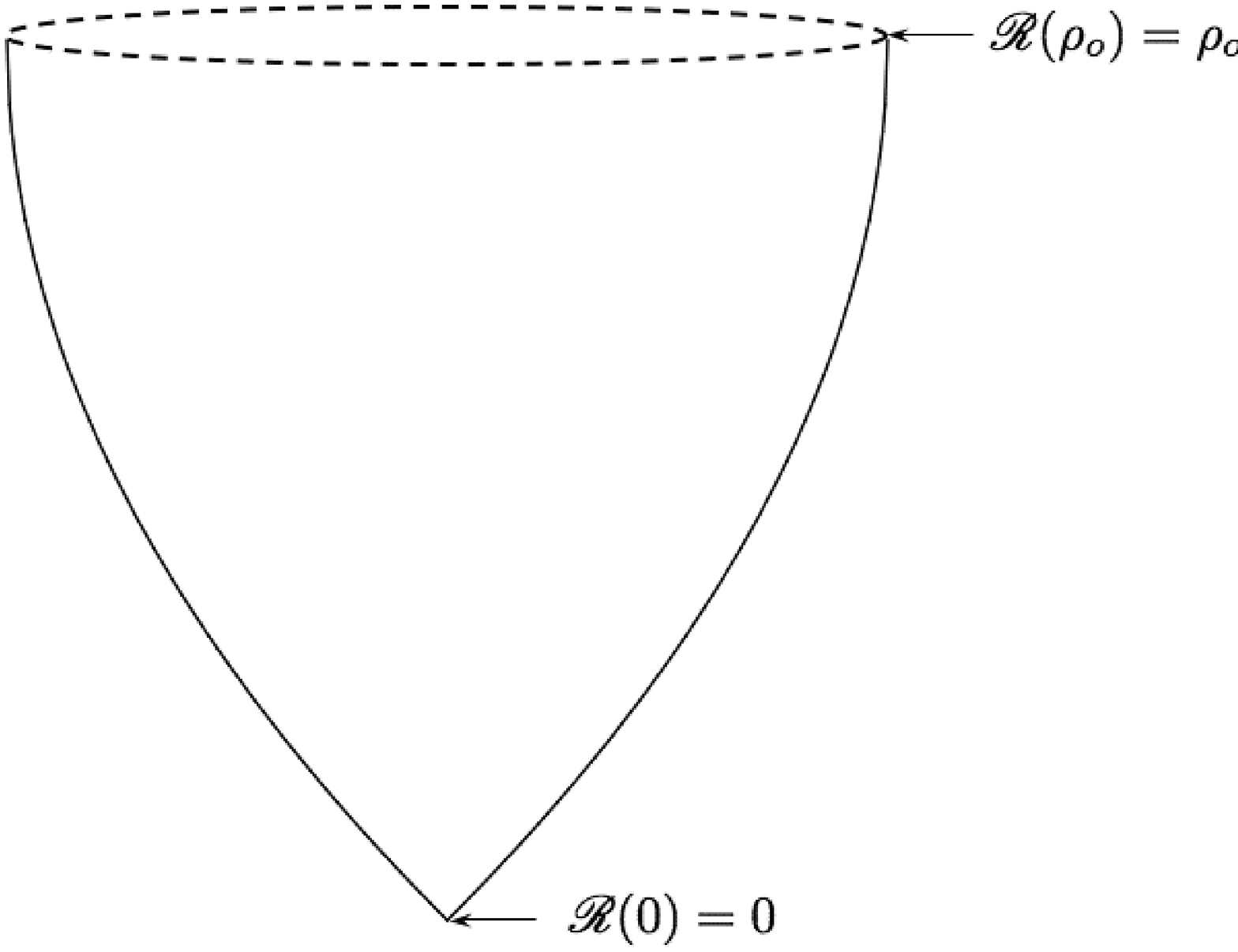}
&\hspace{0.25cm}\includegraphics[height=3.96cm,width=4cm]{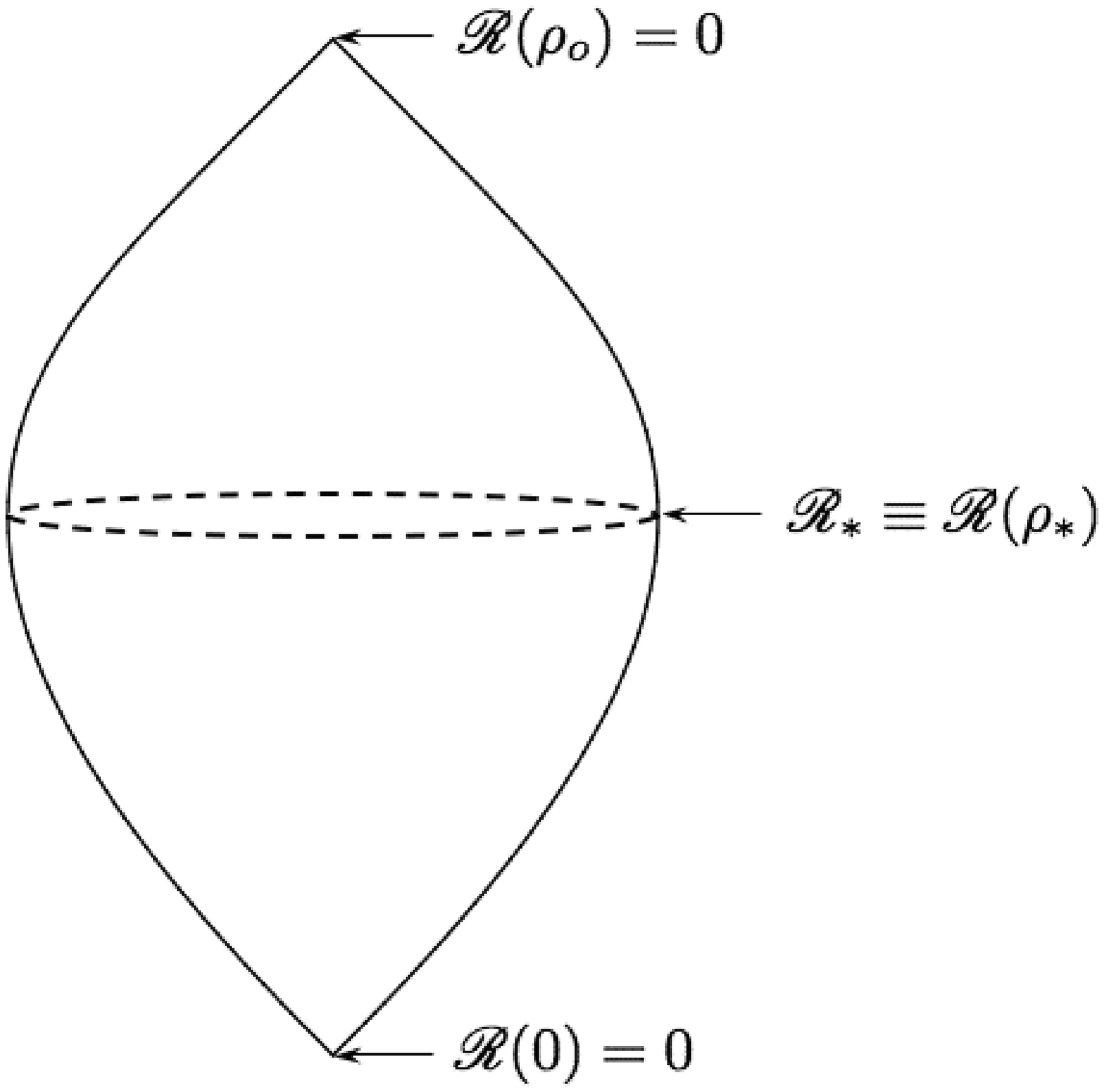}
\nonumber\\
&\hspace{-1.35cm} ({\bf a}) &\hspace{1.25cm}({\bf b})\nonumber
\ea
\vspace{-0.5cm}
\caption{Embedding diagrams for $d=4$. ({\bf a}): Interior region of the Schwarzschild-Tangherlini black hole of $M'=2$ and $\Sigma'=0$,  [see, \eq{st}]. The dashed circle $\mathscr{R}(\rho_o)=\rho_o=r_o$ represents its event horizon and the point $\mathscr{R}(0)=0$ represents its spacelike singularity. ({\bf b}): Fisher universe corresponding to $\mu=1$ and $S=1/2$. The points $\mathscr{R}(\rho_o)=0$ and $\mathscr{R}(0)=0$ represent its spacelike singularities corresponding to the universe's Big Bang and Big Crunch, respectively. The dashed circle of the radius $\mathscr{R}_*$ represents the marginally trapped surface \eq{4.10}. For $S=0$ the diagram is symmetric with respect to the circle. The diagrams are qualitatively generic for other values of $d>4$.}\label{F7}
\end{center} 
\end{figure}

Let us now consider the Fisher universe \eq{3.2} whose 2-dimensional subspace metric is given by  
\be\n{5.8}
ds^2=-\Phi^{\tfrac{\mathstrut 1-S}{d-3}-1}d\rho^2+\rho^2\Phi^{\tfrac{\mathstrut1-S}{d-3}}d\phi^2,
\ee
where $\rho\in(0, \rho_o)$ and $\Phi$ is given by \eq{3.2a}. This surface can be isometrically embedded into a 3-dimensional pseudo-Euclidean space endowed with the following metric:
\be\n{5.9}
dl^2=-d\mathscr{Z}^2+d\mathscr{R}^2+\mathscr{R}^2d\phi^2,
\ee  
Matching the metrics \eq{5.8} and \eq{5.9} we derive the following embedding map:
\ba
\mathscr{R}(\rho)&=&\rho\left[\left(\frac{\rho_o}{\rho}\right)^{d-3}-1\right]^{\tfrac{\mathstrut 1-S}{2(d-3)}}\,,\n{5.10a}\\
\mathscr{Z}(\rho)&=&\rho_o^{\tfrac{\mathstrut d-3}{2}}\int d\rho\hhhhh\frac{[(1+S)^2\rho_o^{d-3}-4S\rho^{d-3}]^{\tfrac{1}{2}}}{2\rho^{\tfrac{\mathstrut 1-S}{2}}(\rho_o^{d-3}-\rho^{d-3})^{1-\tfrac{\mathstrut 1-S}{2(d-3)}}}\,.\non\\\n{5.10b}
\ea
Embeddings of the surfaces corresponding to $S=1$ and $S=1/2$ are presented in Figs.~\ref{F7}({\bf a}) and \ref{F7}({\bf b}), respectively.

We shall discuss the embedding diagrams in the following section.

\section{The Fisher spacetime and the Fisher universe}

So far we were considering the Fisher spacetime and the Fisher universe separately. This approach is based on the fact that the Fisher solution is singular at $r=r_o$ $(\rho=\rho_o)$ and the disconnected parts of the Fisher manifold, which represent the Fisher spacetime and the Fisher universe, seem to not be related to each other. However, we can show that there are certain relations between some geometric quantities of the Fisher spacetime and the Fisher universe. Namely, if we consider expressions \eq{4.5}, \eq{4.5c}, and \eq{4.5e}, we observe that the following relation holds:
\be\n{6.3}
\frac{ \rho_*}{\rho_o}=\frac{r_e}{r_m}=\left(\frac{1+S}{2}\right)^{\tfrac{\mathstrut 1}{d-3}}\,.
\ee
In the limit $S\to1$ we have $\rho_*=\rho_o=r_e=r_m$, where $\rho_o=r_o$ defines the event horizon of the Schwarzschild-Tangherlini black hole which is dual to the Fisher solution. There is an analogous relation between surface areas corresponding to $\rho_*$, $r_e$, $r_m$ and the area $\mathcal{A}_{EH}$ of the black hole event horizon surface $(r=r_o=\rho_o)$ [see Eqs. \eq{4.10},\eq{4.10a},\eq{Rm},\eq{Re}, and \eq{Ae}],
\be\n{6.6}
\frac{\mathcal{A}_{*}}{\mathcal{A}_{EH}}=\frac{\mathcal{A}_e}{\mathcal{A}_m}=\left[\frac{1}{2}(1-S)^{\tfrac{\mathstrut 1-S}{2}}(1+S)^{\tfrac{\mathstrut 1+S}{2}}\right]^{\tfrac{\mathstrut d-2}{d-3}}\,.
\ee
In the limit $S\to1$ we have $\mathcal{A}_{*}=\mathcal{A}_{EH}=\mathcal{A}_e=\mathcal{A}_m$. In addition, in Sec. IV we found that in any dimension $d\geqslant4$ both the areas $\mathcal{A}_{e}$ and $\mathcal{A}_{*}$ calculated for the fixed value of the mass $M=\mu=1$ have minimal values at the same value of $S\approx 0.834$, and for $S\approx 0.611$ they are equal to the horizon surface area of the Schwarzschild-Tangherlini black hole of $M=1$ and $\Sigma=0$ [see Figs.~\ref{F1}({\bf b}) and \ref{F3}({\bf b})]. 

An analysis of the Kretschmann invariant \eq{B2} shows that there is another property which holds for any member of the Fisher family of solutions corresponding to $S\in[0,1)$. Namely, ratio of the Kretschmann invariant to the corresponding squared Ricci scalar \eq{3.7} calculated at $\rho=\rho_*$, $r=r_e$, and $r=r_m$ does not depend on $S$ and M (or $\mu$),
\ba\n{6.7}
\left.\frac{\mathcal{K}}{R^2}\right\vert_{\rho=\rho_*}&=&\frac{2(2d-5)}{(d-2)(d-3)}\,,\\
\left.\frac{\mathcal{K}}{R^2}\right\vert_{r=r_e}&=&\frac{d}{d-2}\,,\\
\left.\frac{\mathcal{K}}{R^2}\right\vert_{r=r_m}&=&\frac{2(2(d-2)^2-1)}{(d-2) (d-3)}\,, 
\ea
where $S\in[0,1)$. Thus, these ratios, as well as $r_o$, are invariants of the duality transformation \eq{2.11} corresponding to $S\in[0,1)$. 

The relations \eq{6.3}, \eq{6.6} may seem ``natural" because both the Fisher spacetime and the Fisher universe originate from the same metric \eq{2.4}. However, such relations may have deeper roots. Our analysis of the Fisher solution yields the following results.  The Schwarzschild-Tangherlini black hole solution belongs to the same theory \eq{2.1}, and it is dual to the Fisher solution. The duality transformation \eq{2.11} maps the exterior region of the Schwarzschild-Tangherlini black hole $r\in(r_o,\infty)$ into the Fisher spacetime $r\in(r_o,\infty)$ and the interior region of the black hole $r\in(0,r_o)$ into the Fisher universe $\rho\in(0,\rho_o)$. Such a map may be visualized with the help of the embedding diagrams presented in Figs.~\ref{F6} and \ref{F7} in Sec. V. Namely, according to expressions \eq{4.10},\eq{areal},\eq{Rm}, and \eq{Re} we have
\be
\mathscr{R}_*|_{S\to1}=R_m|_{S\to1}=R_e|_{S\to1}=R(r_o)=r_o\,.
\ee
This expression implies that in the limit, which corresponds to zero value of the scalar charge, the region between the dashed circle of the radius $R_m$ and the point $R(r_o)=0$ in Fig.~\ref{F6}({\bf b}) maps into the circle of the radius $R(r_o)=r_o$ in Fig.~\ref{F6}({\bf a}), and the region between the dashed circle of the radius $\mathscr{R}_*$ and the point $\mathscr{R}(\rho_o)=0$ in Fig.~\ref{F7}({\bf b}) maps into the circle of the radius $\mathscr{R}(\rho_o)=\rho_o$ in Fig.~\ref{F7}({\bf a}). Both the circles in Figs.~\ref{F6}({\bf a}) and \ref{F7}({\bf a}) represent the event horizon of the Schwarzschild-Tangherlini black hole, i.e., $R(r_o)=r_o=\mathscr{R}(\rho_o)=\rho_o$. Thus, the region of the Fisher spacetime between the $(d-2)$-dimensional sphere of the areal radius $R_m$ and the timelike naked singularity at $r=r_o$ and the region of the Fisher universe between the spacelike naked singularity at $\rho=\rho_o$ and the marginally trapped surface at $\rho=\rho_*$ map into the event horizon of the Schwarzschild-Tangherlini black hole. Note that this is not a one-to-one map. 

\section{Summary and discussion}

In this paper we studied the $d$-dimensional generalization of the Fisher solution, which has a naked curvature singularity that divides the Fisher manifold into two disconnected parts, the Fisher spacetime and the Fisher universe. The $d$-dimensional Schwarzschild-Tangherlini solution and the Fisher solution belong to the same theory \eq{2.1} and are dual to each other. The duality transformation \eq{2.11} maps the exterior region of the Schwarzschild-Tangherlini black hole into the Fisher spacetime, which has a naked timelike singularity, and the interior region of the black hole into the Fisher universe, which is an anisotropic expanding-contracting universe and which has two spacelike singularities representing its Big Bang and Big Crunch. The Big Bang singularity and the singularity of the Fisher spacetime are radially weak in the sense that a 1-dimensional object moving along a timelike radial geodesic can arrive at the singularities intact. These results and the relations between geometric quantities of the Fisher spacetime, the Fisher universe and the Schwarzschild-Tangherlini black hole presented in Sec. VI may suggest the following scenario. The massless scalar field, which according to the results of Sec. III contracts the spacetime in the angular directions, transforms the event horizon of the Schwarzschild-Tangherlini black hole into the naked radially weak disjoint singularities of the Fisher spacetime and the Fisher universe which are ``dual to the horizon." The properties of the Fisher solution presented above may suggest that one could ``join" the Fisher spacetime and the Fisher universe together.  If such a ``junction" is possible, then a 1-dimensional object traveling along a radial geodesic can pass through the timelike naked singularity of the Fisher spacetime and emerge out of the Big Bang singularity into the Fisher universe.

One may think of how to construct a junction between the Fisher spacetime and the Fisher universe. As it was mentioned at the end of Sec. III, one may suggest a $C^0$ local extension of the 2-dimensional $(t,r)$ and $(\tau,\rho)$ spacetime surfaces through the singularities which could provide a junction between the Fisher spacetime and the Fisher universe. However, this does not solve the problem completely, as far as it may provide a 2-dimensional junction only. Thus, one may try to look for other possibilities. For example, in a domain of Planckian curvatures, $\mathcal{K}\sim l^{-4}_{Pl}=c^6/(\hbar G)^2\approx 1.47\times10^{139}m^{-4}$, quantum effects can be dominant and may ``smooth out" curvature singularities. If this is indeed true, then we may expect that the Fisher spacetime and the Fisher universe may be {\em physically} (in the quantum way) joined together. Another way to smooth out the singularities is to consider the Einstein action with higher curvature interactions which are dominant near a spacetime curvature singularity and may remove it. However, there are arguments based on ground state stability which imply that curvature singularities (eternal and timelike) play a useful role as being unphysical \cite{HM}. For example, the timelike singularity of the negative mass Schwarzschild solution, if smoothed out, would give us a negative energy regular solution. As a result, Minkowski spacetime would not be stable. In the case of the Fisher solution, which is a nonvacuum solution, there is a compact region near the singularity (which can be arbitrary small) where the Misner-Sharp energy is negative. However, the energy conditions are not violated. Thus, the singularity of the Fisher spacetime may be ``physical."

How generic can the properties of the Fisher solution be? According to a theorem presented in \cite{Chase} for 4-dimensional spacetime, any static, asymptotically flat solution to Eqs. \eq{2.2} and \eq{2.3} with $\varphi\ne0$ has a singular, simply connected event horizon defined by $k^2=0$, where $-k^2$ is the squared norm of the timelike Killing vector $\delta^a_{\,\,\,t}$ [see, \eq{met}]. The event horizon remains singular if a solution to Eqs. \eq{2.2} and \eq{2.3} with $\varphi\ne0$ is not asymptotically flat. For example, applying the duality transformation \eq{Acd} to a 4-dimensional axisymmetric distorted Schwarzschild black hole discussed in \cite{FS}, we can construct the corresponding axisymmetric distorted Fisher solution. There are other 4-dimensional singular solutions with a massless scalar field which are generalizations of the Fisher solution. These are the Penney solution, which is a generalization of the Reissner-Nordstr\"om solution in the presence of the massless scalar field \cite{Penney} and the Kerr solution with the addition of the massless scalar field \cite{Agn}. These solutions indicate that the massless scalar field transforms the event horizon into a naked singularity. Whether the naked singularity in these solutions is radially weak and the solutions have properties similar to the Fisher solution is an open question. We believe that it is likely to be the case.  

Finally, one can ask if the Fisher solution is physical indeed. This question can be divided into two parts. The first part is whether such a solution can be considered as a result of a gravitational collapse, disproving cosmic censorship conjecture. Spherical gravitational collapse of a massless scalar field (without scalar charge) was studied, e.g., in \cite{Chop,Chris1}. It was found that in some cases naked singularities do appear. However, later it was shown that formation of the naked singularities is an unstable phenomenon \cite{Chris2}. An alternative to gravitational collapse is the existence of primordial singularities (see, e.g., \cite{Mall}). The second part of the question is concerned with the stability of the Fisher solution. To the best of our knowledge this issue is open. The related problem of stability of the negative mass Schwarzschild solution under linearized gravitational perturbations was discussed in \cite{Gib3}. It was found that for a physically preferred boundary conditions corresponding to the perturbations of finite energy the spacetime is stable. A different conclusion concerning to stability of the negative mass Schwarzschild solution had been reached in \cite{GD}. It would be interesting to study the stability of the Fisher spacetime singularity.     

We hope that in the future more can be said about the issues discussed here. 

\begin{acknowledgments}
We would like to thank Mustafa Halilsoy, Don Page, Jutta Kunz, Werner Israel, Rituparno Goswami, and Eric Woolgar for discussions and valuable suggestions. We are grateful to the referee for bringing to our attention details of Fig.~\ref{D4}.
\end{acknowledgments}

\appendix

\section{\label{A} The Einstein and the Klein-Gordon equations}

The Einstein equations \eq{2.2} for a static, spherically symmetric metric of the form
\be\n{A1}
ds^2=-e^Adt^2+e^Bdr^2+e^Cd\Omega^2_{(d-2)}\,,
\ee
where $A,B,C$ are functions of $r$, reduce to
\ba
&&\hspace{-1cm}2A_{,rr}+A_{,r}[A_{,r}-B_{,r}+(d-2)C_{,r}]=0\,,\n{A2a}\\
&&\hspace{-1cm}C_{,r}[2A_{,r}+(d-3)C_{,r}]-4(d-3)e^{B-C}=\frac{4\varphi_{,r}^2}{d-3}\,,\n{A2b}\\
&&\hspace{-1cm}A_{,rr}C_{,r}-C_{,rr}A_{,r}+2(d-3)A_{,r}e^{B-C}=0\,.\n{A2c}
\ea
The Klein-Gordon equation \eq{2.3} for the static, spherically symmetric scalar field $\varphi=\varphi(r)$ is
\be\n{A3}
\left(e^{\frac{1}{2}[A-B+(d-2)C]}\varphi_{,r}\right)_{,r}=0\,.
\ee
Integrating this equation with an appropriate constant of integration we derive
\be\n{A4}
\varphi_{,r}=\frac{4(d-3)\Gamma(\frac{d-1}{2})\Sigma}{(d-2)\pi^{\frac{d-3}{2}}}e^{-\tfrac{\mathstrut 1}{2}[A-B+(d-2)C]}\,.
\ee
A substitution of Eq. \eq{A4} into Eq. \eq{A2b} gives a closed system of equations for the metric functions $A,B,C$.

\section{\label{B} The Riemann tensor and the Kretschmann invariant}
 
The Riemann tensor components for the metric \eq{2.4} defined in a local orthonormal frame are (no summation over $\hal$)
\ba
&&\hspace{-0.4cm}R_{\htt\hr\htt}^{\ \ \ \ \hr}=-\frac{S}{1-S^2}\frac{R}{r_o^{d-3}}\left[2r^{d-3}-(1+S)r_o^{d-3}\right]\,,\n{B1a}\\
&&\hspace{-0.4cm}R_{\htt\hal\htt}^{\ \ \ \ \hal}=-\frac{R_{\htt\hr\htt}^{\hhh \hr}}{d-2}\,,\n{B1b}\\
&&\hspace{-0.4cm}R_{\hr\hal\hr}^{\ \ \ \ \hal}=-\frac{R\left[2Sr^{d-3}-(1+S)r_o^{d-3}\right]}{(1-S^2)(d-2)r_o^{d-3}}\,,\n{B1c}\\
&&\hspace{-0.4cm}R_{\hbe\hal\hbe}^{\ \ \ \ \hal}=\frac{R\left[4Sr^{d-3}-(1+S)^2r_o^{d-3}\right]}{(1-S^2)(d-2)(d-3)r_o^{d-3}}\,,\n{B1d}
\ea
where $R$ is the Ricci scalar \eq{3.7} and the indices $\hal,\hbe=3,...,d$ stand for orthonormal components in the compact dimensions of the $(d-2)$-dimensional round sphere. The corresponding Kretschmann invariant is given by
\ba\n{B2}
\mathcal{K}&\equiv&R_{\ha \hb \hc \hd}R^{\ha\hb\hc\hd}=\frac{2}{(1-S^2)^2}\frac{R^2}{r_o^{2(d-3)}}\left(\frac{d-1}{d-2}\right)\nonumber\\
&\times&\Biggl\{2S^2\left[2r^{d-3}-(1+S)r_0^{d-3}\right]^2\nonumber\\
&+&\frac{2}{(d-1)}\left[2Sr^{d-3}-(1+S)r_0^{d-3}\right]^2\nonumber\\
&+&\frac{1}{(d-1)(d-3)}\left[4Sr^{d-3}-(1+S)^2r_0^{d-3}\right]^2\Biggr\}\,,\non\\
\ea
where $\{\ha,\hb,\hc,\hd\}=\{\htt,\hr;\hal,\hbe=3,...,d\}$.


\begin{thebibliography}{}
\bibitem{Iso} The radial isotropic coordinate $r_i$ of \cite{Xan} is related to our coordinate $r$ through the following transformation:
\be
r_i^{d-3}=\frac{r^{d-3}}{4}\left(1+\sqrt{F}\right)^2\,,\hhh F=1-\left(\frac{r_o}{r}\right)^{d-3}\,.\non
\ee
\bibitem{complex} One can show that for an appropriate discrete set of the parameters $M$ and $\Sigma$ the metric \eq{2.4} can be real valued for $r\in(0,r_o)$. Here we shall not consider such a restrictive choice of the parameters.
\bibitem{RN} Let us note that for a Reisner-Nordstr\"om spacetime behavior of the corresponding functions $M(R)$ and $U(R)$ is qualitatively the same. However, the Misner-Sharp energy is negative in the region behind the Cauchy horizon, and maximum of $U(R)\geqslant1/2$, were the equality sign stands for the extremal Reisner-Nordstr\"om spacetime.
\bibitem{TOP} Here, and in what follows by topology of a spacetime singularity, we mean topology of ideal points of a spacetime which represent the singularity (see, e.g., \cite{Haw,GKR}).
\end{thebibliography}
\end{document}